\documentclass[aps,prl,twocolumn,superscriptaddress,nofootinbib,floatfix,10pt]{revtex4}

\usepackage{amsmath,amssymb,amsfonts}
\usepackage{bm}
\usepackage{graphicx}
\usepackage{xcolor}
\usepackage[colorlinks=true,citecolor=blue,linkcolor=blue,urlcolor=blue]{hyperref}
\newcommand{\supp}{\operatorname{supp}}
\newcommand{\Ker}{\operatorname{Ker}}
\newcommand{\Span}{\operatorname{span}}
\newcommand{\ket}[1]{|#1\rangle}
\newcommand{\bra}[1]{\langle #1|}

\begin{document}

\title{Boundary Geometry Turns Entanglement into Steering}

\author{Yu-Xuan Zhang}
\affiliation{School of Physics, Nankai University, Tianjin 300071, People's Republic of China}

\author{Jing-Ling Chen}
\email{chenjl@nankai.edu.cn}
\affiliation{Theoretical Physics Division, Chern Institute of Mathematics, Nankai University, Tianjin 300071, People's Republic of China}

\date{\today}

\begin{abstract}
Quantum entanglement does not necessarily imply Einstein-Podolsky-Rosen steering. We identify a \emph{boundary mechanism} that closes this gap when an entangled state meets the boundary of the trusted state space in a nondegenerate way. The mechanism is local: a projective assemblage that approaches a boundary contact with first-order tangential coherence but only second-order inward defect cannot be reproduced by any finite-measure local-hidden-state model. In finite dimensions this yields a support-kernel criterion for an arbitrary fixed steering cut: if a trusted conditional state is rank-deficient and its first-order block couples its support to its kernel, then the state is both NPT (negative partial transpose) and projectively steerable. The two-qubit product-null sector provides the minimal transparent realization. A product vector in the kernel gives the pure boundary contact, and a single coherence entry simultaneously serves as the tangential displacement, the NPT minor, and the steering obstruction. Hence, on this boundary, entanglement collapses to two-way projective steering --- including the genuinely mixed rank-three branch and the rank-two case as a low-dimensional instance. We also provide a filtered standard form, an explicit rank-three Cholesky parametrization, and a compact boundary certificate.
\end{abstract}

\maketitle

\emph{Introduction.}--- Einstein-Podolsky-Rosen (EPR) steering is a form of quantum nonlocality in which measurements by an untrusted party remotely prepare conditional states for a trusted party in a way that cannot be explained by a local-hidden-state (LHS) model \cite{EPR1935,Schrodinger1935,Wiseman2007,Jones2007,Uola2020,Cavalcanti2016}.  It occupies an intermediate position in the hierarchy \({\rm Bell\ nonlocality}\Rightarrow {\rm steering}\Rightarrow {\rm entanglement}\), while the reverse implications fail in general \cite{Bell1964,Werner1989,Horodecki2009,Wiseman2007,Jones2007,Uola2020}.  In particular, steering is directional, and entanglement alone is not usually a steering certificate \cite{Werner1989,Bowles2014}.  The question addressed here is when the boundary geometry of the state itself leaves no room for an LHS simulation.  This viewpoint complements steering inequalities, semidefinite-programming methods, measurement incompatibility, and ellipsoid techniques \cite{Pusey2013,Skrzypczyk2014,Nguyen2019,Quintino2014,Uola2014,Uola2015}.

\begin{figure}[t]
\centering
\includegraphics[width=0.9\columnwidth]{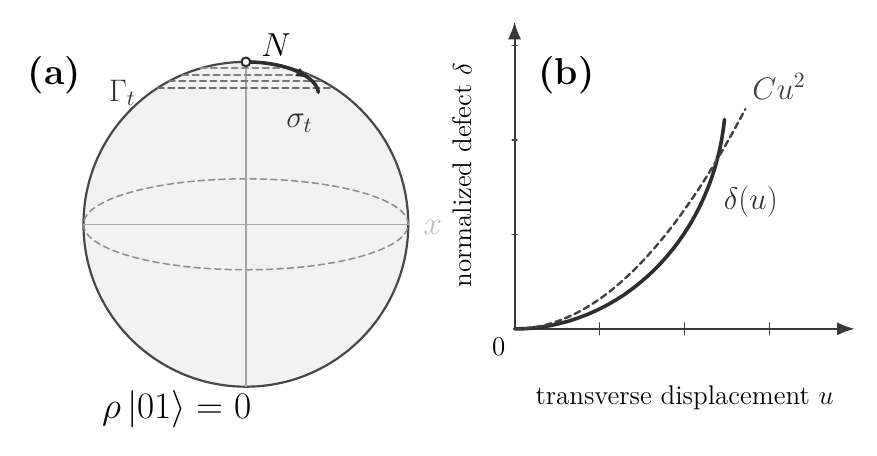}
\caption{Boundary contact and quadratic inward defect.  (a) Product-null boundary contact at \(N\), with shrinking caps \(\Gamma_t\).  (b) The normalized inward defect \(\delta(u)\) is bounded by a quadratic curve \(Cu^2\) for a representative product-null family.}
\label{fig:boundary-contact}
\end{figure}

The mechanism is easiest to visualize in its minimal two-qubit realization.  A product vector in the kernel is not merely a rank defect: it forces a trusted conditional state to acquire a zero direction.  For a trusted qubit this is a pure boundary contact with the Bloch sphere, and in the steering-ellipsoid picture it is a tangency between the steering ellipsoid and the Bloch sphere \cite{Jevtic2014,Jevtic2015,NguyenVu2016,SongBakerWiseman2023}.  The contact becomes nondegenerate when nearby conditional states leave the boundary with first-order tangential coherence while positivity fills the orthogonal direction only to second order.  The resulting boundary scaling is incompatible with any fixed finite hidden-state measure.  A hidden atom at the contact point is harmless, since it has no tangential component; after the contact face is removed, the punctured neighborhoods that could supply the tangential motion shrink to the empty set for a finite measure.  This local geometry is illustrated in Fig.~\ref{fig:boundary-contact}.

The same mechanism has a finite-dimensional support-kernel form.  For an arbitrary fixed cut \(X|Y\), where \(Y\) may be high-dimensional or multipartite, a rank-deficient trusted conditional state defines a boundary face of the positive cone.  If the first-order block of the state couples the support of this boundary state to its kernel, then the same two-dimensional compression gives the boundary scaling obstruction and, at the same time, a negative partial-transpose minor.  Thus the two-qubit product-null sector should be read as the minimal explicit realization of a support-kernel boundary mechanism, not as the source of the mechanism itself.

On the two-qubit product-null boundary this general idea becomes especially sharp.  After local unitaries the product-null vector can be written as \(\ket{01}\), and a single matrix entry \(h_{02}=\bra{00}\rho\ket{11}\) controls both entanglement and steering.  It is the leading tangential coherence at the boundary contact and also the off-diagonal entry of the NPT minor.  Hence the usual hierarchy collapses on this boundary stratum: entanglement is equivalent to two-way projective steering.  We further identify a steering-safe filtered standard form, give an explicit Cholesky parametrization of the genuinely mixed rank-three branch, and record the high-dimensional support-kernel criterion as the general form of the same boundary principle.

For two-qubit product-null states, the general support-kernel obstruction becomes exact on the corresponding boundary stratum.  By local unitaries one may place the product-null vector at \(\ket{01}\).  The tangency is then controlled by a single tangential coherence,
\(h_{02}=\bra{00}\rho\ket{11}.\)
This same matrix element has two roles.  First, it is the first-order coefficient that makes Bob's conditional states leave the contact point tangentially and hence triggers the boundary-contact scaling obstruction.  Second, it gives a negative principal minor of the partial transpose.  By the Peres-Horodecki criterion, PPT is equivalent to separability for two qubits \cite{Peres1996,Horodecki1996}.  Therefore, on the product-null boundary, the same tangential coherence detects entanglement and certifies projective steering.

This yields a sharp collapse of the usual hierarchy on the product-null boundary:
\({\rm entanglement} \Longleftrightarrow {\rm two\mbox{-}way\ projective\ steering}.\)
As a consequence, every entangled two-qubit rank-two state is two-way projectively steerable, because the two-dimensional kernel of a rank-two two-qubit state always contains a product vector.  This rank-two consequence recovers the previously established rank-two steering theorem from a different viewpoint \cite{ZhangChen2025Fundamental}.  The present proof is different: it identifies the product-null tangency as the minimal realization of the local boundary-contact scaling obstruction.  The same mechanism also covers the rank-three branch whose null space is spanned by a product vector.

This boundary result has a simple operational reading: once a candidate product-null direction is fixed, a boundary population test checks the contact, and the single tangential coherence directly determines the NPT minor. Thus the mechanism turns a boundary NPT witness into a projective-steering certificate without requiring an independent steering inequality.

We also provide a steering-safe filtered standard form, an explicit Cholesky parametrization of the rank-three branch, and a compact boundary certificate. Finally, the support-kernel mechanism extends to arbitrary finite dimensions and multipartite trusted systems, where it remains a sufficient condition for NPT entanglement and projective steering across a fixed cut. The entanglement-steering gap therefore depends not only on the amount of entanglement but crucially on how the state meets the boundary of the trusted state space.

%The boundary result also has a simple operational reading.  Once a candidate product-null direction is fixed, the boundary population checks the contact, while a single tangential coherence determines the partial-transpose minor.  Thus, on the product-null boundary, the boundary mechanism converts a boundary NPT witness into a projective-steering certificate, rather than requiring an independent steering-inequality optimization.
%
%Finally, the method is not tied to the special global geometry of the Bloch ball.  The Bloch-sphere tangency is the two-dimensional shadow of a more general support-kernel coupling at the boundary of the trusted density-matrix cone.  For an arbitrary steering cut \(X|Y\), with \(Y\) possibly high-dimensional or multipartite, choose two orthonormal vectors \(\ket{\alpha_0},\ket{\alpha_1}\in{\cal H}_X\) and define
%\(A=\bra{\alpha_0}\rho\ket{\alpha_0}, \qquad B=\bra{\alpha_0}\rho\ket{\alpha_1}.\)
%If \(A\) is nonzero and rank deficient, and if
%\(P_{\supp A}BP_{\Ker A}\ne0,\)
%then a two-dimensional trusted compression reproduces the same boundary-contact scaling obstruction.  The same support-kernel coherence also gives a negative principal minor of the partial transpose across the cut.  Hence the state is NPT across \(X|Y\) and projectively steerable from \(X\) to \(Y\).  This is the finite-dimensional support-kernel form of the boundary mechanism; the product-null two-qubit case is its minimal sharp specialization.

\emph{Boundary-contact scaling obstruction.}--- We first isolate the local obstruction behind the proof.  At this stage no special form of the bipartite state is used; the relevant object is only a two-dimensional trusted slice near a boundary contact.  Let \(\sigma_t\), \(0<t<t_0\), be unnormalized qubit conditional states.  Choose the trusted basis so that the normalized states approach the pure boundary point \(\ket0\bra0\) as \(t\to0\), and write
\(\sigma_t=\left(\begin{smallmatrix}a_t&b_t\\ \bar b_t&d_t\end{smallmatrix}\right)\).
Assume that, for all sufficiently small \(t>0\),
\begin{eqnarray}
|b_t|\ge lt,
\qquad
d_t\le ct^2,
\label{eq:boundary-scaling-condition}
\end{eqnarray}
with constants \(l>0\) and \(c<\infty\).  The entry \(b_t\) is the support-kernel coherence in this two-dimensional slice, while \(d_t\) is the weight in the orthogonal direction.  Thus Eq.~\eqref{eq:boundary-scaling-condition} describes first-order tangential motion along the boundary and only second-order inward defect.

The tension is simple.  Hidden states away from the contact point can carry transverse components, but they also carry a non-negligible inward defect.  Hidden states very close to the contact point have small inward defect, but their transverse components are small as well.  To produce a first-order transverse displacement while keeping only a second-order inward cost, an LHS model would have to keep a fixed amount of hidden weight inside caps whose radius shrinks linearly with \(t\).  No LHS model with a fixed finite hidden-state measure can do this; in particular, the usual normalized LHS models are included.  In Bloch form,
\(\sigma_t=(m_tI+\vec R_t\cdot\vec\sigma)/2\),
write
\(\vec R_{t,\perp}=(R_{t,x},R_{t,y})\)
for the component tangent to the Bloch sphere at the limiting north pole.  Then
\(|\vec R_{t,\perp}|=2|b_t|\) and \(m_t-R_{t,z}=2d_t\).
Hence Eq.~\eqref{eq:boundary-scaling-condition} implies
\begin{eqnarray}
|\vec R_{t,\perp}|\ge Lt,
\qquad
m_t-R_{t,z}\le Ct^2
\label{eq:bloch-boundary-scaling}
\end{eqnarray}
for some constants \(L>0\) and \(C<\infty\).  After normalization, the variables in Fig.~\ref{fig:boundary-contact}(b) are \(u_t=|\vec R_{t,\perp}|/m_t\) and \(\delta_t=(m_t-R_{t,z})/m_t=2d_t/m_t\); near a nonzero boundary contact, this is the scaling \(u_t=O(t)\), \(\delta_t=O(t^2)\).

Suppose, to the contrary, that an LHS representation exists.  After pushing the hidden states forward to their Bloch vectors, there are a finite measure \(\nu\) on the Bloch ball and response functions \(0\le f_t(\vec r)\le1\) such that
\begin{eqnarray}
m_t
=
\int f_t(\vec r)\,d\nu(\vec r),
\qquad
\vec R_t
=
\int f_t(\vec r)\vec r\,d\nu(\vec r).
\label{eq:lhs-bloch-representation}
\end{eqnarray}
Write
\(\vec r=(x,y,z), \qquad \vec r_\perp=(x,y), \qquad w(\vec r)=1-z .\)
For every point in the Bloch ball,
\begin{eqnarray}
|\vec r_\perp|^2
=
x^2+y^2
\le
1-z^2
\le
2(1-z)
=
2w(\vec r).
\label{eq:cap-geometry}
\end{eqnarray}
Moreover, Eq.~\eqref{eq:lhs-bloch-representation} gives
\(\int f_t(\vec r)w(\vec r)\,d\nu(\vec r)=m_t-R_{t,z}\le Ct^2\).

Choose \(K>4C/L\), and let
\(\Gamma_t=\{\vec r:\ 1-z\le K^2t^2/2\}\)
be a cap around the north pole \(N=(0,0,1)\).  On the complement \(\Gamma_t^c\), one has \(w(\vec r)>K^2t^2/2\).  Combining this with Eq.~\eqref{eq:cap-geometry} gives
\(|\vec r_\perp|\le 2w(\vec r)/(Kt)\) for \(\vec r\in\Gamma_t^c\).
Therefore the transverse contribution from the complement satisfies
\begin{eqnarray}
\left|
\int_{\Gamma_t^c}
f_t(\vec r)\vec r_\perp\,d\nu(\vec r)
\right|
&\le&
\int_{\Gamma_t^c}
f_t(\vec r)|\vec r_\perp|\,d\nu(\vec r)
\nonumber\\
&\le&
\frac{2}{Kt}
\int_{\Gamma_t^c}
f_t(\vec r)w(\vec r)\,d\nu(\vec r)
\nonumber\\
&\le&
\frac{2C}{K}t
<
\frac{L}{2}t.
\label{eq:cap-complement-bound}
\end{eqnarray}
This is the key estimate.  The cap complement may contain hidden states with sizable transverse components, but using them would create too much inward defect.  Thus any LHS explanation of the transverse motion is forced into the shrinking cap itself.

Since the total transverse displacement satisfies \(|\vec R_{t,\perp}|\ge Lt\), Eq.~\eqref{eq:cap-complement-bound} implies that the cap itself must contribute at least order \(t\):
\(\left|\int_{\Gamma_t} f_t(\vec r)\vec r_\perp\,d\nu(\vec r)\right|\ge Lt/2\).
On the other hand, inside \(\Gamma_t\), Eq.~\eqref{eq:cap-geometry} gives
\(|\vec r_\perp|\le Kt\).
The pole \(N\) itself has \(\vec r_\perp=0\), and hence does not contribute to the transverse integral.  Thus an atom of the hidden measure at \(N\) is allowed, but it cannot help produce the required tangential displacement.  Therefore the transverse lower bound forces a uniform positive amount of hidden measure away from the pole:
\begin{eqnarray}
\nu(\Gamma_t\setminus\{N\})
\ge
\frac{L}{2K}.
\label{eq:positive-cap-measure}
\end{eqnarray}

This is impossible for a finite measure.  Along any sequence \(t_n\downarrow0\), the punctured caps
\(\Gamma_{t_n}\setminus\{N\}\)
decrease to the empty set, and finite measures are continuous from above.  Hence
\(\nu(\Gamma_{t_n}\setminus\{N\})\to0\),
contradicting Eq.~\eqref{eq:positive-cap-measure}.  Thus no finite-measure LHS model can reproduce a family satisfying Eq.~\eqref{eq:boundary-scaling-condition}.  The Supplemental Material gives the same cap estimate in full detail~\cite{SM}.

The conclusion should therefore be read as a local boundary statement rather than as a finite-setting steering inequality.  It says that no finite-measure LHS model can sustain first-order support-kernel coherence at a boundary contact while paying only a second-order inward cost.

\emph{Product-null realization and low-rank consequences.}--- We now specialize the support-kernel boundary mechanism to its minimal two-qubit product-null realization.  Suppose that \(\ket{\alpha}\otimes\ket{\beta}\in\Ker\rho\).  By local unitaries we take this vector to be \(\ket{01}\).  Since \(\rho\ge0\), the whole \(\ket{01}\)-row and column vanish, and
\begin{eqnarray}
\rho
=
\frac{1}{\operatorname{tr}H}
\left(
\begin{array}{cccc}
h_{00}&0&h_{01}&h_{02}\\
0&0&0&0\\
\bar h_{01}&0&h_{11}&h_{12}\\
\bar h_{02}&0&\bar h_{12}&h_{22}
\end{array}
\right),
\qquad
H\ge0.
\label{eq:product-null-standard-form}
\end{eqnarray}
Conversely, any nonzero-probability pure boundary contact of Bob's conditional state forces a corresponding product vector into \(\Ker\rho\); hence Eq.~\eqref{eq:product-null-standard-form} is the local normal form of the two-qubit pure-contact realization.  Alice's outcome \(\ket0\bra0\) prepares on Bob's side a conditional state supported on \(\ket0\), i.e., a point on the Bloch sphere.  In the steering-ellipsoid picture this is exactly a tangency between Bob's ellipsoid and the Bloch sphere.

The contact is nondegenerate exactly when \(h_{02}\ne0\).  For \(\ket{\xi_t}=(\ket0+t\ket1)/\sqrt{1+t^2}\) and \(\sigma_t=\bra{\xi_t}\rho\ket{\xi_t}\), the relevant Bob-side entries are
\begin{eqnarray}
b_t
=
\frac{t h_{02}+t^2h_{12}}
{(\operatorname{tr}H)(1+t^2)},
\qquad
d_t
=
\frac{t^2h_{22}}
{(\operatorname{tr}H)(1+t^2)}.
\label{eq:product-null-bd}
\end{eqnarray}
Thus \(h_{02}\ne0\) gives first-order support-kernel coherence, \(|b_t|\ge lt\), while the inward defect remains second order, \(d_t\le ct^2\), for all sufficiently small \(t>0\).  The boundary-contact obstruction then gives projective steering from \(A\) to \(B\).

The same entry controls entanglement.  In \(\rho^{\Gamma_B}\), the principal block on \(\Span\{\ket{01},\ket{10}\}\) is \((\operatorname{tr}H)^{-1}\left(\begin{smallmatrix}0&h_{02}\\ \bar h_{02}&h_{11}\end{smallmatrix}\right)\), whose determinant is negative whenever \(h_{02}\ne0\).  Conversely, \(h_{02}=0\) makes the product-null standard family PPT; see the Supplemental Material~\cite{SM}.  Since PPT is equivalent to separability for two qubits \cite{Peres1996,Horodecki1996}, \(h_{02}\) is simultaneously the support-kernel coherence, the leading tangential displacement, and the NPT minor.  Hence, on the product-null boundary,
\begin{eqnarray}
h_{02}\ne0
\quad\Longleftrightarrow\quad
\rho\in{\rm Ent}
\quad\Longleftrightarrow\quad
\rho\in{\rm PS}_{A\to B}.
\label{eq:product-null-oneway-equivalence}
\end{eqnarray}
Applying the same argument after exchanging the two parties gives the reversed direction.  Therefore
\begin{eqnarray}
\rho\in{\rm Ent}
\quad\Longleftrightarrow\quad
\rho\in{\rm TPS}
\label{eq:product-null-ent-tps}
\end{eqnarray}
on the two-qubit product-null boundary, where \({\rm TPS}\) denotes two-way projective steering.

The standard matrix form should be understood up to steering-safe local transformations.  For \(A\to B\), Bob is trusted, so an invertible filter on Bob transforms hidden states and can be absorbed into the hidden-state measure; Alice is untrusted, so on Alice's side we use local unitaries to keep projective measurements projective.  Equivalently, the invariant condition is the existence of local vectors \(\alpha,\beta\) with \(\rho(\alpha\otimes\beta)=0\) and a nonzero support-kernel block \(M_{\alpha,\alpha^\perp}\beta\ne0\), where \(M_{\alpha,\alpha^\perp}:=(\bra{\alpha}\otimes I)\rho(\ket{\alpha^\perp}\otimes I)\).  The details are given in the Supplemental Material~\cite{SM}.

The genuinely mixed branch is rank three.  If \(H>0\), then \(\Ker\rho=\Span\{\ket{01}\}\).  Write \(H=LL^\dagger\) with
\begin{eqnarray}
L=
\left(
\begin{array}{ccc}
a&0&0\\
x&b&0\\
z&y&c
\end{array}
\right),
\qquad a,b,c>0.
\end{eqnarray}
Then \(h_{02}=a\bar z\), so \(z\ne0\) is exactly the entanglement and two-way projective-steering condition on this rank-three branch.  Equivalently, in the spectral parametrization on the three-dimensional support, the same condition is the nonvanishing of the support coherence \(C_{03}\), where \(C_{03}=\sum_k\nu_kV_{1k}\bar V_{3k}\) and \(V\in U(3)/U(1)^3\); an Euler-angle chart is given in the Supplemental Material~\cite{SM}.  Lower ranks follow as limits.  For rank two, the kernel is two-dimensional and therefore contains a product vector; every entangled rank-two two-qubit state is two-way projectively steerable, recovering Ref.~\cite{ZhangChen2025Fundamental} from the boundary-contact viewpoint.  Rank-one entangled states are included as the pure-state case.

\emph{Boundary certificate.}--- In the exact product-null case, the boundary mechanism turns a population test and a single support-kernel coherence measurement into a steering certificate.  Let \(\ket\alpha\otimes\ket\beta\) be a candidate product-null vector.  Since \(\rho\ge0\), \(p_0:=\bra{\alpha,\beta}\rho\ket{\alpha,\beta}=0\) implies \(\rho\ket{\alpha,\beta}=0\) and thus identifies the boundary contact.  Choose local bases with \(\ket\alpha=\ket0_A\) and \(\ket\beta=\ket1_B\).  The tangential (support-kernel) coherence is \(C_{\rm tan}:=\bra{00}\rho\ket{11}\).  At the product-null boundary, \(C_{\rm tan}\ne0\) is equivalent to NPT entanglement and, via Eq.~\eqref{eq:product-null-ent-tps}, to two-way projective steering.

A robust NPT witness is given by the boundary partial-transpose minor
\begin{eqnarray}
W_{\rm bd}=|\rho_{00,11}|^2-\rho_{01,01}\rho_{10,10}. \label{eq:boundary-witness}
\end{eqnarray}
If \(W_{\rm bd}>0\), then \(\rho^{\Gamma_B}\not\ge0\).  In the ideal tangency limit \(\rho_{01,01}=0\), this reduces to \(|C_{\rm tan}|^2>0\).  With finite experimental accuracy, one measures the small product-null population \(p_0=\rho_{01,01}\) and evaluates the full minor; a positive \(W_{\rm bd}\) indicates a robust boundary NPT witness.  The steering conclusion holds whenever the product-null contact is exact or guaranteed by the source.

In the standard Pauli basis,
\begin{eqnarray}
C_{\rm tan}= \frac14\bigl[\langle XX\rangle-\langle YY\rangle - i(\langle XY\rangle+\langle YX\rangle)\bigr]. \label{eq:coherence-pauli}
\end{eqnarray}
For real-valued implementations the imaginary part vanishes, so only the two settings \(XX\) and \(YY\) are needed, together with the computational-basis population test.  The protocol thus resembles a boundary entanglement-witness measurement rather than a conventional steering experiment.  The nontrivial step is provided by the boundary mechanism: at an exact product-null contact, the same local NPT datum becomes a projective-steering certificate, consistent with the steering-incompatibility correspondence \cite{Quintino2014,Uola2014,Uola2015}.

A simple preparable family is \(\rho^{(\Phi)}_{p,q,r}=p\Phi_+ + q\Phi_- + r\ket{10}\bra{10}\), where \(\Phi_\pm=\ket{\Phi^\pm}\bra{\Phi^\pm}\), \(\ket{\Phi^\pm}=(\ket{00}\pm\ket{11})/\sqrt2\), and \(p+q+r=1\).  Here \(\ket{01}\) is product-null and \(C_{\rm tan}=(p-q)/2\); steering reduces to the imbalance \(p\ne q\).  When \(p=q\), the Bell coherences cancel and the tangency becomes degenerate; moving away from this plane in either direction creates the nondegenerate boundary contact detected by the witness.

\emph{Higher-dimensional and multipartite support-kernel form.}--- The product-null construction is the most economical realization of a more general support-kernel boundary mechanism.  What matters is not the global shape of the Bloch ball, but the presence of a zero direction in a trusted conditional state and a first-order block that couples this zero direction to the occupied part.

Consider an arbitrary steering cut \(X|Y\), where \(Y\) may be high-dimensional or multipartite.  Choose two orthonormal vectors \(\ket{\alpha_0},\ket{\alpha_1}\in{\cal H}_X\), and write the corresponding two-dimensional \(X\)-block of \(\rho\) as \(\left(\begin{smallmatrix}A&B\\ B^\dagger&D\end{smallmatrix}\right)_X\), where \(A,B,D\) act on \({\cal H}_Y\).  The projective vector \(\ket{\xi_t}=(\ket{\alpha_0}+t\ket{\alpha_1})/\sqrt{1+t^2}\) prepares \(\sigma_t=[A+t(B+B^\dagger)+t^2D]/(1+t^2)\).  Assume that \(A\ne0\) has a nontrivial kernel and that \(B\) connects the support of \(A\) to its kernel,
\(P_{\supp A}BP_{\Ker A}\ne0\).
Equivalently, there are unit vectors \(\ket\phi\in\supp A\) and \(\ket\beta\in\Ker A\) such that \(v:=\bra\phi B\ket\beta\ne0\).

Restricting the trusted system to \({\cal K}=\Span\{\ket\phi,\ket\beta\}\) gives the same local scaling as in the qubit proof.  Since \(\rho\ge0\) and \(\ket\beta\in\Ker A\), the vector \(\ket{\alpha_0,\beta}\) is annihilated by \(\rho\), and hence \(B^\dagger\ket\beta=0\).  Thus the compressed conditional states have first-order support-kernel coherence \(tv+O(t^2)\) and only second-order population in the kernel direction.  The boundary-contact obstruction rules out an LHS model for the compressed family and therefore for the original projective assemblage.  The same matrix element gives an NPT minor of \(\rho^{\Gamma_Y}\) on \(\Span\{\ket{\alpha_0,\beta},\ket{\alpha_1,\phi}\}\).  Consequently,
\begin{eqnarray}
\left.
\begin{array}{c}
A\ne0,\quad \Ker A\ne\{0\},\\
P_{\supp A}BP_{\Ker A}\ne0
\end{array}
\right\}
\Longrightarrow
\left\{
\begin{array}{c}
\rho^{\Gamma_Y}\not\ge0,\\
\rho\in{\rm PS}_{X\to Y}.
\end{array}
\right.
\label{eq:highdim-criterion}
\end{eqnarray}
This is the finite-dimensional support-kernel form of the boundary mechanism for a fixed steering cut.  It is not a general equivalence between entanglement and steering; its content is structural.  Whenever a boundary conditional state has a zero direction and the first-order change couples that zero direction to the occupied part, the same local datum gives both the NPT minor and the projective-steering obstruction.

There is, however, an important equivalence class in arbitrary finite dimensions.  Suppose the relevant block is a pure-contact standard block supported on the two-dimensional untrusted sector above, with \(A=a\ket\phi\bra\phi\) and \(a>0\).  Positivity forces \(B=\ket\phi\bra w\).  If the support-kernel part of \(w\) vanishes, then \(B=\kappa\ket\phi\bra\phi\), the Schur complement gives \(D-(|\kappa|^2/a)\ket\phi\bra\phi\ge0\), and the block is separable across \(X|Y\).  If the support-kernel part is nonzero, Eq.~\eqref{eq:highdim-criterion} gives NPT entanglement and projective steering.  Thus, within this pure-contact standard family, entanglement, NPT, support-kernel coherence, and projective steering are equivalent for the fixed cut.  The two-qubit product-null theorem is the minimal explicit realization of this pure-contact equivalence, while Eq.~\eqref{eq:highdim-criterion} remains a sufficient test for mixed boundary contacts.

The criterion only uses a two-dimensional slice of the trusted system, so the remaining Hilbert space may be large, structured, or multipartite.  If \(Y\) is multipartite, the statement concerns the fixed cut \(X|Y\), with the systems inside \(Y\) treated jointly.  It does not by itself imply genuine multipartite steering, which would require excluding hybrid LHS models across different cuts.

\emph{Conclusion.}--- We have identified a boundary-contact mechanism by which entanglement can force projective steering.  The mechanism is local: a rank-deficient trusted conditional state supplies a boundary face, and a first-order support-kernel coherence drives the conditional states tangentially along that face while positivity supplies the kernel population only at second order.  A finite-measure LHS model cannot reproduce this scaling, because it would have to keep a uniform positive selected mass in punctured neighborhoods that shrink to the contact point after the contact face is removed.

For two qubits this mechanism has a minimal and complete product-null realization.  The product-null vector gives the pure boundary contact, and the single entry \(h_{02}\) is simultaneously the support-kernel coherence, the tangential displacement, the off-diagonal element of an NPT minor, and the trigger for the boundary-contact obstruction.  Consequently, on the product-null boundary, \({\rm Ent}\Longleftrightarrow {\rm TPS}\).  The rank-three product-null branch gives the genuinely mixed realization of this collapse, while the rank-two theorem follows as a low-rank consequence because a two-dimensional two-qubit kernel contains a product vector.

The broader lesson is that the entanglement-steering gap is not controlled only by the amount of entanglement.  It also depends on how the state meets the boundary of the trusted state space.  Product-null two-qubit states provide the cleanest illustration, while the same support-kernel form applies to arbitrary finite-dimensional fixed cuts and to multipartite trusted systems treated jointly.  This perspective turns a local boundary NPT minor into a projective-steering certificate whenever the appropriate boundary contact is exact or source-guaranteed, and it suggests a systematic way to look for further families where the usual hierarchy collapses.

\begin{acknowledgments}
This work is supported by the Quantum Science and Technology--National Science and Technology Major Project (Grant No. 2024ZD0301000), and the National Natural Science Foundation of China (Grant No. 12275136).
\end{acknowledgments}

\emph{Note added.}--- During the completion of this manuscript, we became aware of independent related work by Song and Bae on pure steered states and two-qubit EPR steering from the steering-ellipsoid perspective~\cite{SongBae2026}.  The two works were developed independently and emphasize complementary aspects of the same boundary-contact phenomenon.

\end{document}

\title{Supplemental Material for ``Boundary Geometry Turns Entanglement into Steering''}

\author{Yu-Xuan Zhang}
\affiliation{School of Physics, Nankai University, Tianjin 300071, People's Republic of China}

\author{Jing-Ling Chen}
\email{chenjl@nankai.edu.cn}
\affiliation{Theoretical Physics Division, Chern Institute of Mathematics, Nankai University, Tianjin 300071, People's Republic of
	 	China}

\date{\today}

%\maketitle

\maketitle

\section{The boundary-contact scaling obstruction}

In this section we establish a basic obstruction which will be used later.  Consider a one-parameter family of unnormalized conditional states on Bob's side,
\begin{eqnarray}
\{\sigma_t\}_{0<t<t_0}.
\end{eqnarray}
Here \(t\) is a small positive parameter labeling this family of Bob conditional states.  We assume that, as \(t\to0\), the normalized direction of this family approaches a pure state on the boundary of the Bloch ball.  By choosing a local basis on Bob's side, we may assume without loss of generality that this boundary pure state is
\begin{eqnarray}
\ket0\bra0,
\end{eqnarray}
namely the north pole of the Bloch ball.

In this Bob basis, write
\begin{eqnarray}
\sigma_t=
\left(
\begin{array}{cc}
a_t&b_t\\
\bar b_t&d_t
\end{array}
\right),
\qquad
\sigma_t\ge0.
\label{eq:sigma-t-matrix}
\end{eqnarray}
Equivalently, write
\begin{eqnarray}
\sigma_t=
\frac{1}{2}
\left(
m_t I+\vec R_t\cdot\vec\sigma
\right),
\qquad
m_t=\tr\sigma_t,
\qquad
\vec R_t=(R_{t,x},R_{t,y},R_{t,z}).
\label{eq:sigma-t-bloch}
\end{eqnarray}
Comparing the two representations gives
\begin{eqnarray}
m_t=a_t+d_t,
\qquad
R_{t,x}=2{\rm Re}\,b_t,
\qquad
R_{t,y}=-2{\rm Im}\,b_t,
\qquad
R_{t,z}=a_t-d_t.
\end{eqnarray}
Hence
\begin{eqnarray}
|\vec R_{t,\perp}|=2|b_t|,
\qquad
m_t-R_{t,z}=2d_t,
\label{eq:matrix-bloch-defect}
\end{eqnarray}
where
\begin{eqnarray}
\vec R_{t,\perp}=(R_{t,x},R_{t,y}).
\end{eqnarray}
For the normalized Bloch vector, the corresponding transverse displacement and inward defect are
\begin{eqnarray}
u_t=\frac{|\vec R_{t,\perp}|}{m_t},
\qquad
\delta_t=\frac{m_t-R_{t,z}}{m_t}=\frac{2d_t}{m_t}.
\end{eqnarray}
Thus, in this fixed Bob basis, if
\begin{eqnarray}
|b_t|=O(t),
\qquad
d_t=O(t^2),
\end{eqnarray}
then the family of Bob conditional states has a first-order transverse displacement near the boundary, while its inward defect from the boundary is only of second order.

We now show that this behavior cannot be simulated by a finite LHS measure.  This includes the usual normalized LHS models, whose hidden-state measure is a probability measure.  Suppose that there is a usual LHS representation
\begin{eqnarray}
\sigma_t=
\int_\Lambda f_t(\lambda)\tau_\lambda\,d\mu(\lambda),
\qquad
0\le f_t(\lambda)\le1,
\label{eq:usual-lhs}
\end{eqnarray}
where \(\mu\) is a finite measure on the hidden-variable space \(\Lambda\), and each \(\tau_\lambda\) is a normalized qubit state.  Write each hidden state in Bloch form as
\begin{eqnarray}
\tau_\lambda=
\frac{1}{2}
\left(
I+\vec r(\lambda)\cdot\vec\sigma
\right).
\end{eqnarray}
Let \(\nu\) be the finite Borel measure on the Bloch ball \(\mathbb B^3\) induced by \(\mu\) under the map
\begin{eqnarray}
\lambda\mapsto \vec r(\lambda).
\end{eqnarray}
Equivalently, for every Borel set \(E\subset\mathbb B^3\),
\begin{eqnarray}
\nu(E)
=
\mu\left(
\{\lambda:\vec r(\lambda)\in E\}
\right).
\end{eqnarray}
For each \(t\), define a positive measure \(\alpha_t\) on \(\mathbb B^3\) by
\begin{eqnarray}
\alpha_t(E)
=
\int_{\{\lambda:\vec r(\lambda)\in E\}}
f_t(\lambda)\,d\mu(\lambda).
\end{eqnarray}
Since \(0\le f_t(\lambda)\le1\), one has
\begin{eqnarray}
0\le \alpha_t(E)\le \nu(E).
\end{eqnarray}
Therefore \(\alpha_t\) is absolutely continuous with respect to \(\nu\).  By the Radon--Nikodym theorem, there exists a function, still denoted by \(f_t(\vec r)\), such that
\begin{eqnarray}
d\alpha_t(\vec r)
=
f_t(\vec r)\,d\nu(\vec r),
\qquad
0\le f_t(\vec r)\le1.
\end{eqnarray}
Thus the LHS representation can equivalently be written on the Bloch ball as
\begin{eqnarray}
m_t=
\int_{\mathbb B^3}
f_t(\vec r)\,d\nu(\vec r),
\qquad
\vec R_t=
\int_{\mathbb B^3}
f_t(\vec r)\vec r\,d\nu(\vec r).
\label{eq:bloch-lhs-representation}
\end{eqnarray}
The important point is that the measure \(\nu\) is independent of \(t\); only the response function \(f_t\) may depend on \(t\).

Assume now that there exist constants \(L>0\) and \(C<\infty\) such that, for all sufficiently small \(t>0\),
\begin{eqnarray}
|\vec R_{t,\perp}|\ge Lt,
\qquad
m_t-R_{t,z}\le Ct^2.
\label{eq:boundary-contact-scaling-assumption}
\end{eqnarray}
We shall derive a contradiction.

Let
\begin{eqnarray}
\vec r=(x,y,z),
\qquad
\vec r_\perp=(x,y),
\qquad
w(\vec r)=1-z.
\end{eqnarray}
Since \(\vec r\in\mathbb B^3\),
\begin{eqnarray}
|\vec r_\perp|^2
=
x^2+y^2
\le
1-z^2
=
(1-z)(1+z)
\le
2(1-z)
=
2w(\vec r).
\label{eq:bloch-ball-transverse-bound}
\end{eqnarray}
On the other hand, by the Bloch-ball LHS representation,
\begin{eqnarray}
\int_{\mathbb B^3}
f_t(\vec r)w(\vec r)\,d\nu(\vec r)
&=&
\int_{\mathbb B^3}
f_t(\vec r)(1-z)\,d\nu(\vec r)
\nonumber\\
&=&
m_t-R_{t,z}.
\end{eqnarray}
Hence \Eq{eq:boundary-contact-scaling-assumption} gives
\begin{eqnarray}
\int_{\mathbb B^3}
f_t(\vec r)w(\vec r)\,d\nu(\vec r)
\le
Ct^2.
\label{eq:u-small}
\end{eqnarray}

We now split the Bloch ball into a small cap around the north pole and its complement.  Choose a constant \(K\) such that
\begin{eqnarray}
K>\frac{4C}{L}.
\end{eqnarray}
Let
\begin{eqnarray}
N=(0,0,1),
\end{eqnarray}
and define
\begin{eqnarray}
\Gamma_t=
\left\{
\vec r\in\mathbb B^3:
1-z\le\frac{K^2t^2}{2}
\right\}.
\label{eq:shrinking-cap}
\end{eqnarray}
This is a cap of height \(O(t^2)\) around the north pole.

We first estimate the transverse contribution from the complement \(\Gamma_t^c\).  On \(\Gamma_t^c\),
\begin{eqnarray}
w(\vec r)=1-z>\frac{K^2t^2}{2}.
\end{eqnarray}
Together with \Eq{eq:bloch-ball-transverse-bound}, this implies
\begin{eqnarray}
|\vec r_\perp|
\le
\sqrt{2w(\vec r)}
\le
\frac{2w(\vec r)}{Kt}.
\label{eq:outside-cap-bound}
\end{eqnarray}
Therefore
\begin{eqnarray}
\left|
\int_{\Gamma_t^c}
f_t(\vec r)\vec r_\perp\,d\nu(\vec r)
\right|
&\le&
\int_{\Gamma_t^c}
f_t(\vec r)|\vec r_\perp|\,d\nu(\vec r)
\nonumber\\
&\le&
\frac{2}{Kt}
\int_{\Gamma_t^c}
f_t(\vec r)w(\vec r)\,d\nu(\vec r)
\nonumber\\
&\le&
\frac{2}{Kt}
\int_{\mathbb B^3}
f_t(\vec r)w(\vec r)\,d\nu(\vec r)
\nonumber\\
&\le&
\frac{2C}{K}t
<
\frac{L}{2}t.
\label{eq:outside-small}
\end{eqnarray}
Thus the transverse contribution from the complement of the cap has norm less than \(Lt/2\).

Since the total transverse vector satisfies
\begin{eqnarray}
|\vec R_{t,\perp}|\ge Lt,
\end{eqnarray}
the cap must contribute at least one half of it:
\begin{eqnarray}
\left|
\int_{\Gamma_t}
f_t(\vec r)\vec r_\perp\,d\nu(\vec r)
\right|
\ge
\frac{L}{2}t.
\label{eq:inside-large}
\end{eqnarray}
On the other hand, inside \(\Gamma_t\), we have
\begin{eqnarray}
1-z\le\frac{K^2t^2}{2}.
\end{eqnarray}
Using \Eq{eq:bloch-ball-transverse-bound} again gives
\begin{eqnarray}
|\vec r_\perp|\le Kt.
\end{eqnarray}
The north pole \(N\) itself satisfies \(\vec r_\perp=0\), and hence it does not affect the transverse integral.  Thus an atom of the hidden measure at \(N\) is allowed, but it cannot help produce the required transverse displacement.  Therefore,
\begin{eqnarray}
\left|
\int_{\Gamma_t}
f_t(\vec r)\vec r_\perp\,d\nu(\vec r)
\right|
&=&
\left|
\int_{\Gamma_t\setminus\{N\}}
f_t(\vec r)\vec r_\perp\,d\nu(\vec r)
\right|
\nonumber\\
&\le&
Kt\,\nu(\Gamma_t\setminus\{N\}).
\label{eq:cap-upper}
\end{eqnarray}
Combining \Eq{eq:inside-large} and \Eq{eq:cap-upper}, we obtain
\begin{eqnarray}
\nu(\Gamma_t\setminus\{N\})
\ge
\frac{L}{2K}
\label{eq:cap-mass-lower-bound}
\end{eqnarray}
for all sufficiently small \(t>0\).

This contradicts the finiteness of \(\nu\).  Indeed, take any sequence \(t_n\downarrow0\).  Then the sets
\begin{eqnarray}
\Gamma_{t_n}\setminus\{N\}
\end{eqnarray}
are decreasing, and their intersection is empty, since
\begin{eqnarray}
\bigcap_{n=1}^{\infty}\Gamma_{t_n}
=
\{N\}.
\end{eqnarray}
By continuity from above for finite measures,
\begin{eqnarray}
\nu(\Gamma_{t_n}\setminus\{N\})
\longrightarrow0.
\end{eqnarray}
This contradicts the uniform positive lower bound \Eq{eq:cap-mass-lower-bound}.  Therefore no fixed finite hidden-state measure can simulate the boundary-contact scaling behavior \Eq{eq:boundary-contact-scaling-assumption}; the contradiction comes from the positive measure forced into punctured caps shrinking to the contact point.

We shall use the following matrix criterion.  If a projective assemblage contains a one-parameter family of Bob conditional states
\begin{eqnarray}
\sigma_t=
\left(
\begin{array}{cc}
a_t&b_t\\
\bar b_t&d_t
\end{array}
\right),
\end{eqnarray}
and the Bob basis is chosen so that the boundary pure state approached as \(t\to0\) is the Bloch north pole, then the existence of constants \(l>0\) and \(c<\infty\) with
\begin{eqnarray}
|b_t|\ge lt,
\qquad
d_t\le ct^2
\end{eqnarray}
for all sufficiently small \(t>0\) rules out any LHS model with a finite hidden-state measure.

\section{The standard product-null family}

We now consider the standard product-null family of two-qubit states.  This family is not only a convenient construction; it is the local normal form of any nonzero-probability contact between a steered Bob state and the boundary of the Bloch ball.

To see this, let
\begin{eqnarray}
P_\alpha=\ket{\alpha}\bra{\alpha}
\end{eqnarray}
be a rank-one projector on Alice's side, and let
\begin{eqnarray}
\sigma_\alpha
=
\operatorname{tr}_A[(P_\alpha\otimes I_B)\rho]
\end{eqnarray}
be the corresponding unnormalized Bob conditional state.  Suppose that this outcome occurs with nonzero probability and that the normalized Bob conditional state is pure:
\begin{eqnarray}
\operatorname{tr}\sigma_\alpha>0,
\qquad
\frac{\sigma_\alpha}{\operatorname{tr}\sigma_\alpha}
=
\ket{\phi}\bra{\phi}.
\end{eqnarray}
Then
\begin{eqnarray}
\sigma_\alpha\ket{\phi^\perp}=0.
\end{eqnarray}
Hence
\begin{eqnarray}
0
=
\bra{\phi^\perp}\sigma_\alpha\ket{\phi^\perp}
=
\bra{\alpha,\phi^\perp}\rho\ket{\alpha,\phi^\perp}.
\end{eqnarray}
Since \(\rho\ge0\), this implies
\begin{eqnarray}
\rho\ket{\alpha,\phi^\perp}=0.
\end{eqnarray}
Thus every nonzero-probability contact with a pure point of Bob's Bloch sphere forces a product vector into the kernel of \(\rho\).

Conversely, if
\begin{eqnarray}
\ket{\alpha}\otimes\ket{\beta}\in{\rm Ker}\,\rho
\end{eqnarray}
and
\begin{eqnarray}
\operatorname{tr}\sigma_\alpha>0,
\end{eqnarray}
then the Bob conditional state \(\sigma_\alpha\) has a zero direction \(\ket{\beta}\).  Since Bob is a qubit, a nonzero positive operator with a zero direction is rank one.  Therefore the normalized Bob conditional state is pure.  We therefore have the local equivalence
\begin{eqnarray}
\text{nonzero-probability contact with Bob's Bloch sphere}
\Longleftrightarrow
\text{a corresponding product vector in }{\rm Ker}\,\rho.
\end{eqnarray}

After local unitary changes of basis, we may take
\begin{eqnarray}
\ket{\alpha}=\ket0_A,
\qquad
\ket{\beta}=\ket1_B.
\end{eqnarray}
Thus the product-null vector becomes
\begin{eqnarray}
\ket{01}.
\end{eqnarray}
In the computational basis
\begin{eqnarray}
\ket{00},\ket{01},\ket{10},\ket{11},
\end{eqnarray}
the positivity of \(\rho\) then forces the whole \(\ket{01}\)-row and \(\ket{01}\)-column to vanish.  Hence \(\rho\) has the standard product-null form
\begin{eqnarray}
\rho=
\frac{1}{\operatorname{tr} H}
\left(
\begin{array}{cccc}
h_{00}&0&h_{01}&h_{02}\\
0&0&0&0\\
\bar h_{01}&0&h_{11}&h_{12}\\
\bar h_{02}&0&\bar h_{12}&h_{22}
\end{array}
\right),
\label{eq:standard-product-null-family}
\end{eqnarray}
where
\begin{eqnarray}
H=
\left(
\begin{array}{ccc}
h_{00}&h_{01}&h_{02}\\
\bar h_{01}&h_{11}&h_{12}\\
\bar h_{02}&\bar h_{12}&h_{22}
\end{array}
\right)\ge0,
\qquad
\operatorname{tr}H>0.
\label{eq:H-block-standard-family}
\end{eqnarray}
Here \(H\) is the block of \(\rho\) on the three-dimensional subspace
\begin{eqnarray}
{\rm span}\{\ket{00},\ket{10},\ket{11}\}.
\end{eqnarray}
The rank of \(\rho\) is equal to the rank of \(H\).  In particular, \(H>0\) gives the rank-three product-null case, while \({\rm rank}\,H=2\) gives the rank-two case.  In all cases,
\begin{eqnarray}
\rho\ket{01}=0.
\end{eqnarray}

We next consider a one-parameter family of binary projective measurements on Alice's side.  For small positive \(t\), define
\begin{eqnarray}
\ket{\xi_t}
=
\frac{\ket0+t\ket1}{\sqrt{1+t^2}},
\end{eqnarray}
and
\begin{eqnarray}
P_t=\ket{\xi_t}\bra{\xi_t},
\qquad
Q_t=I_A-P_t.
\end{eqnarray}
Then \(\{P_t,Q_t\}\) is a binary rank-one projective measurement on Alice's side.  We only need the unnormalized Bob conditional state corresponding to the outcome \(P_t\):
\begin{eqnarray}
\sigma_t
=
\operatorname{tr}_A[
(P_t\otimes I_B)\rho
].
\end{eqnarray}
A direct calculation gives
\begin{eqnarray}
\sigma_t
=
\frac{1}{(\operatorname{tr} H)(1+t^2)}
\left(
\begin{array}{cc}
h_{00}+2t{\rm Re}\,h_{01}+t^2h_{11}
&
t h_{02}+t^2h_{12}
\\
t\bar h_{02}+t^2\bar h_{12}
&
t^2h_{22}
\end{array}
\right).
\label{eq:standard-sigma-t}
\end{eqnarray}
Equivalently, write \(\rho\) in Alice block form,
\begin{eqnarray}
\rho=
\frac{1}{\operatorname{tr} H}
\left(
\begin{array}{cc}
A&B\\
B^\dagger&D
\end{array}
\right)_A,
\end{eqnarray}
where
\begin{eqnarray}
A=
\left(
\begin{array}{cc}
h_{00}&0\\
0&0
\end{array}
\right),
\quad
B=
\left(
\begin{array}{cc}
h_{01}&h_{02}\\
0&0
\end{array}
\right),
\quad
D=
\left(
\begin{array}{cc}
h_{11}&h_{12}\\
\bar h_{12}&h_{22}
\end{array}
\right).
\end{eqnarray}
Then
\begin{eqnarray}
\sigma_t
=
\frac{A+tB+tB^\dagger+t^2D}
{(\operatorname{tr} H)(1+t^2)}.
\end{eqnarray}

At \(t=0\), the conditional state is
\begin{eqnarray}
\sigma_0=
\frac{h_{00}}{\operatorname{tr} H}\ket0\bra0.
\end{eqnarray}
Thus, whenever \(h_{00}>0\), the normalized Bob conditional state touches the boundary of the Bloch ball at the pure state \(\ket0\bra0\).  In the present argument, the stronger condition \(h_{02}\ne0\) automatically implies this contact.  Indeed, positivity of the principal block of \(H\) on the first and third coordinates gives
\begin{eqnarray}
|h_{02}|^2\le h_{00}h_{22}.
\end{eqnarray}
Hence
\begin{eqnarray}
h_{02}\ne0
\quad\Longrightarrow\quad
h_{00}>0,
\qquad
h_{22}>0.
\end{eqnarray}
Therefore the family \(\sigma_t\) reaches the Bloch-sphere boundary at \(t=0\).  The additional role of \(h_{02}\) is not the boundary contact itself, but the existence of a nonzero first-order tangential displacement away from the contact point.

We now inspect the matrix entries of \(\sigma_t\).  Its off-diagonal entry is
\begin{eqnarray}
b_t=
\frac{t h_{02}+t^2h_{12}}
{(\operatorname{tr} H)(1+t^2)}.
\end{eqnarray}
If
\begin{eqnarray}
h_{02}\ne0,
\end{eqnarray}
then, for all sufficiently small \(t>0\),
\begin{eqnarray}
|t h_{02}+t^2h_{12}|
=
t|h_{02}+t h_{12}|
\ge
\frac{|h_{02}|}{2}t.
\end{eqnarray}
Since \(1+t^2\) is bounded near \(t=0\), there exists a constant \(l>0\) such that
\begin{eqnarray}
|b_t|\ge lt
\end{eqnarray}
for all sufficiently small \(t>0\).  On the other hand, the lower-right entry of \(\sigma_t\) is
\begin{eqnarray}
d_t=
\frac{t^2h_{22}}
{(\operatorname{tr} H)(1+t^2)}.
\end{eqnarray}
Since \(H\ge0\), \(h_{22}\ge0\), and hence there exists a constant \(c<\infty\) such that
\begin{eqnarray}
d_t\le ct^2.
\end{eqnarray}
Therefore, whenever \(h_{02}\ne0\), the Bob conditional states satisfy
\begin{eqnarray}
|b_t|\ge lt,
\qquad
d_t\le ct^2.
\end{eqnarray}

This is precisely the local boundary-scaling pattern at a pure boundary point of the Bloch ball.  The entry \(b_t\) is the support-kernel, or tangential, coherence relative to the limiting pure state \(\ket0\bra0\), while \(d_t\) measures the inward weight in the orthogonal direction \(\ket1\).  Thus the product-null vector gives the boundary contact, and the coherence entry \(h_{02}\) gives the first-order support-kernel motion.

By the boundary-contact scaling obstruction, this projective assemblage cannot admit an LHS model with a finite hidden-state measure.  Indeed, if the whole binary projective assemblage had an LHS model, then the subfamily corresponding to the outcome \(P_t\) would also have the required LHS representation.  Since this subfamily already violates the finite-measure obstruction, the whole projective assemblage is not LHS.  Thus
\begin{eqnarray}
h_{02}\ne0
\quad\Longrightarrow\quad
\rho \ {\rm is\ projectively\ steerable\ from}\ A\ {\rm to}\ B.
\end{eqnarray}

We next relate the same condition to entanglement within the standard family.  The partial transpose with respect to Bob is
\begin{eqnarray}
\rho^{\Gamma_B}
=
\frac{1}{\operatorname{tr} H}
\left(
\begin{array}{cccc}
h_{00}&0&h_{01}&0\\
0&0&h_{02}&0\\
\bar h_{01}&\bar h_{02}&h_{11}&\bar h_{12}\\
0&0&h_{12}&h_{22}
\end{array}
\right).
\label{eq:standard-partial-transpose}
\end{eqnarray}
If \(h_{02}\ne0\), then the principal submatrix of \(\rho^{\Gamma_B}\) on
\begin{eqnarray}
{\rm span}\{\ket{01},\ket{10}\}
\end{eqnarray}
is
\begin{eqnarray}
\frac{1}{\operatorname{tr} H}
\left(
\begin{array}{cc}
0&h_{02}\\
\bar h_{02}&h_{11}
\end{array}
\right).
\end{eqnarray}
Its determinant is
\begin{eqnarray}
-\frac{|h_{02}|^2}{(\operatorname{tr} H)^2}<0.
\end{eqnarray}
Therefore
\begin{eqnarray}
\rho^{\Gamma_B}\not\ge0.
\end{eqnarray}
The state \(\rho\) is NPT, and hence entangled.

Conversely, suppose that
\begin{eqnarray}
h_{02}=0.
\end{eqnarray}
Then
\begin{eqnarray}
\rho=
\frac{1}{\operatorname{tr} H}
\left(
\begin{array}{cccc}
h_{00}&0&h_{01}&0\\
0&0&0&0\\
\bar h_{01}&0&h_{11}&h_{12}\\
0&0&\bar h_{12}&h_{22}
\end{array}
\right),
\end{eqnarray}
where
\begin{eqnarray}
H=
\left(
\begin{array}{ccc}
h_{00}&h_{01}&0\\
\bar h_{01}&h_{11}&h_{12}\\
0&\bar h_{12}&h_{22}
\end{array}
\right)\ge0.
\end{eqnarray}
The partial transpose becomes
\begin{eqnarray}
\rho^{\Gamma_B}
=
\frac{1}{\operatorname{tr} H}
\left(
\begin{array}{cccc}
h_{00}&0&h_{01}&0\\
0&0&0&0\\
\bar h_{01}&0&h_{11}&\bar h_{12}\\
0&0&h_{12}&h_{22}
\end{array}
\right).
\end{eqnarray}
Thus it is enough to check positivity of the nonzero \(3\times3\) block
\begin{eqnarray}
H'=
\left(
\begin{array}{ccc}
h_{00}&h_{01}&0\\
\bar h_{01}&h_{11}&\bar h_{12}\\
0&h_{12}&h_{22}
\end{array}
\right).
\end{eqnarray}
If \(h_{12}=0\), then \(H'=H\).  If \(h_{12}\ne0\), write
\begin{eqnarray}
h_{12}=|h_{12}|e^{i\theta}
\end{eqnarray}
and take
\begin{eqnarray}
D={\rm diag}(1,1,e^{2i\theta}).
\end{eqnarray}
In both cases, one has
\begin{eqnarray}
H'=DHD^\dagger
\end{eqnarray}
for a suitable diagonal unitary \(D\).  Hence \(H'\ge0\), and therefore
\begin{eqnarray}
\rho^{\Gamma_B}\ge0.
\end{eqnarray}
For two qubits, PPT is equivalent to separability.  Hence \(\rho\) is separable.

Thus, within the standard family,
\begin{eqnarray}
\rho\ {\rm is\ entangled}
\quad\Longleftrightarrow\quad
\rho\ {\rm is\ NPT}
\quad\Longleftrightarrow\quad
h_{02}\ne0.
\end{eqnarray}
Combining this equivalence with the boundary-scaling calculation gives
\begin{eqnarray}
\rho\ {\rm is\ entangled}
\quad\Longrightarrow\quad
\rho \ {\rm is\ projectively\ steerable\ from}\ A\ {\rm to}\ B.
\end{eqnarray}

The same nonzero entry also gives the swapped direction.  The zero vector is
\begin{eqnarray}
\ket0_A\otimes\ket1_B.
\end{eqnarray}
For the \(A\to B\) direction, the relevant tangential coherence entry is
\begin{eqnarray}
\bra{00}\rho\ket{11}
=
\frac{h_{02}}{\operatorname{tr} H}.
\end{eqnarray}
After exchanging \(A\) and \(B\), the same zero vector is viewed as
\begin{eqnarray}
\ket1_B\otimes\ket0_A.
\end{eqnarray}
In the \(B\to A\) direction, the corresponding tangential coherence entry is
\begin{eqnarray}
\bra{11}\rho\ket{00}
=
\frac{\bar h_{02}}{\operatorname{tr} H}.
\end{eqnarray}
Thus \(h_{02}\ne0\) triggers the same boundary-scaling mechanism in the swapped direction.  Consequently, within the standard family,
\begin{eqnarray}
\rho\ {\rm is\ entangled}
\quad\Longrightarrow\quad
\rho \ {\rm is\ two\mbox{-}way\ projectively\ steerable}.
\end{eqnarray}

The preceding discussion also explains why the standard family is the natural local form of the mechanism.  Any nonzero-probability steering contact with a pure point of Bob's Bloch sphere forces a product vector into the kernel of \(\rho\).  After local unitaries, this gives the product-null form \(\rho\ket{01}=0\).  A nonzero tangential coherence entry \(h_{02}\) is then exactly the condition that the steered states leave the contact point with a first-order transverse component.  Thus the standard product-null family is not merely a sufficient construction; it is the local normal form of the boundary-contact scaling mechanism.

\section{The filtered standard class and the rank-two case}

The standard form studied above is basis-dependent.  What is invariant is not the particular matrix pattern itself, but the possibility of bringing a state into that pattern by operations that are safe for projective steering.

For the \(A\to B\) direction, Bob is the trusted party.  Hence an invertible filter on Bob's side is allowed.  Alice, on the other hand, is the untrusted party; on her side we only use local unitaries, since a non-unitary filter on Alice's side would in general pull projective measurements back to POVMs.  Thus we consider states which, after a local unitary change of basis on Alice and an invertible filter on Bob, take the standard form
\begin{eqnarray}
\omega=
\left(
\begin{array}{cccc}
a&0&u&v\\
0&0&0&0\\
\bar u&0&c&e\\
\bar v&0&\bar e&d
\end{array}
\right),
\qquad
v\ne0.
\label{eq:filtered-standard-form}
\end{eqnarray}
Equivalently,
\begin{eqnarray}
\omega\ket{01}=0,
\qquad
\bra{00}\omega\ket{11}\ne0.
\end{eqnarray}
By the previous section, such a state triggers the boundary-contact scaling obstruction and is therefore projectively steerable from \(A\) to \(B\).

Let us recall why the Bob-side filtering step is safe.  Suppose that \(\rho\) has an \(A\to B\) projective LHS model.  If
\begin{eqnarray}
\omega=
\frac{(I_A\otimes G_B)\rho(I_A\otimes G_B^\dagger)}
{\tr[(I_A\otimes G_B)\rho(I_A\otimes G_B^\dagger)]}
\end{eqnarray}
with \(G_B\) invertible, then the hidden states are transformed as
\begin{eqnarray}
\tau_\lambda
\longmapsto
\frac{G_B\tau_\lambda G_B^\dagger}
{\tr(G_B\tau_\lambda G_B^\dagger)},
\end{eqnarray}
and the positive factor
\begin{eqnarray}
\frac{\tr(G_B\tau_\lambda G_B^\dagger)}
{\tr[(I_A\otimes G_B)\rho(I_A\otimes G_B^\dagger)]}
\end{eqnarray}
is absorbed into the hidden measure.  The response functions on Alice's side are unchanged.  Thus a projective LHS model for \(\rho\) would imply one for \(\omega\).  Consequently, if such a filtered state \(\omega\) is projectively steerable, then the original state \(\rho\) is also projectively steerable.

The same class can be recognized without first choosing the standard basis.  Assume that there are local vectors \(\alpha\in\mathbb C_A^2\) and \(\beta\in\mathbb C_B^2\) such that
\begin{eqnarray}
\rho(\alpha\otimes\beta)=0.
\label{eq:product-zero-vector}
\end{eqnarray}
Choose a unit vector \(\alpha_\perp\) orthogonal to \(\alpha\), and define the Bob-side operator
\begin{eqnarray}
M_{\alpha,\alpha_\perp}
=
(\bra{\alpha}\otimes I_B)\rho(\ket{\alpha_\perp}\otimes I_B).
\label{eq:M-alpha-alpha-perp}
\end{eqnarray}
If
\begin{eqnarray}
M_{\alpha,\alpha_\perp}\beta\ne0,
\label{eq:tangential-coherence-condition}
\end{eqnarray}
then \(\rho\) belongs to the filtered standard class.

Indeed, set
\begin{eqnarray}
\eta=
M_{\alpha,\alpha_\perp}\beta.
\end{eqnarray}
Since \(\rho\ge0\) and \(\rho(\alpha\otimes\beta)=0\), one also has
\begin{eqnarray}
(\bra{\alpha}\otimes\bra{\beta})\rho=0.
\end{eqnarray}
Therefore
\begin{eqnarray}
\langle\beta|\eta\rangle
&=&
(\bra{\alpha}\otimes\bra{\beta})
\rho
(\ket{\alpha_\perp}\otimes\ket{\beta})
\nonumber\\
&=&0.
\end{eqnarray}
Thus \(\eta\perp\beta\).  If \(\eta\ne0\), then \(\eta\) and \(\beta\) form a basis of Bob's two-dimensional space.

After a local unitary change of basis on Alice, we may assume
\begin{eqnarray}
\alpha=\ket0_A,
\qquad
\alpha_\perp=\ket1_A.
\end{eqnarray}
Choose an invertible Bob filter \(G_B\) such that
\begin{eqnarray}
G_B^\dagger\ket1_B=\beta,
\qquad
G_B^\dagger\ket0_B=\eta.
\end{eqnarray}
This is possible because \(\eta\) and \(\beta\) are linearly independent.  Define
\begin{eqnarray}
\omega=
\frac{(I_A\otimes G_B)\rho(I_A\otimes G_B^\dagger)}
{\tr[(I_A\otimes G_B)\rho(I_A\otimes G_B^\dagger)]}.
\end{eqnarray}
Then
\begin{eqnarray}
(I_A\otimes G_B^\dagger)\ket{01}
=
\ket0_A\otimes\beta
=
\alpha\otimes\beta,
\end{eqnarray}
and hence
\begin{eqnarray}
\omega\ket{01}=0.
\end{eqnarray}
In the computational basis, this means that the \(\ket{01}\)-column of \(\omega\) is zero; since \(\omega\) is Hermitian, the \(\ket{01}\)-row is zero as well.  Thus \(\omega\) has the form
\begin{eqnarray}
\omega=
\left(
\begin{array}{cccc}
a&0&u&v\\
0&0&0&0\\
\bar u&0&c&e\\
\bar v&0&\bar e&d
\end{array}
\right).
\end{eqnarray}
Moreover,
\begin{eqnarray}
\bra{00}\omega\ket{11}
&=&
\frac{
(\bra{\alpha}\otimes\bra{\eta})
\rho
(\ket{\alpha_\perp}\otimes\ket{\beta})
}{
\tr[(I_A\otimes G_B)\rho(I_A\otimes G_B^\dagger)]
}
\nonumber\\
&=&
\frac{
\langle\eta|M_{\alpha,\alpha_\perp}\beta\rangle
}{
\tr[(I_A\otimes G_B)\rho(I_A\otimes G_B^\dagger)]
}
\nonumber\\
&=&
\frac{\|\eta\|^2}
{\tr[(I_A\otimes G_B)\rho(I_A\otimes G_B^\dagger)]}
\ne0.
\end{eqnarray}
Thus the filtered state is exactly in the standard scaling form.

Conversely, if a state is already in the standard form \Eq{eq:filtered-standard-form}, then taking
\begin{eqnarray}
\alpha=\ket0_A,
\qquad
\alpha_\perp=\ket1_A,
\qquad
\beta=\ket1_B
\end{eqnarray}
gives
\begin{eqnarray}
\rho(\alpha\otimes\beta)=0,
\qquad
M_{\alpha,\alpha_\perp}\beta\ne0.
\end{eqnarray}
Therefore the filtered standard class is equivalently characterized by the two conditions
\begin{eqnarray}
\rho(\alpha\otimes\beta)=0,
\qquad
M_{\alpha,\alpha_\perp}\beta\ne0.
\label{eq:coordinate-free-standard-class}
\end{eqnarray}

This characterization also makes the symmetry between the two steering directions transparent.  For the reversed direction \(B\to A\), the same argument is applied after exchanging the roles of the two parties; the allowed invertible filter is then on the trusted side \(A\).

To see the corresponding tangential coherence explicitly, let
\begin{eqnarray}
\eta=M_{\alpha,\alpha_\perp}\beta\ne0.
\end{eqnarray}
As shown above,
\begin{eqnarray}
\eta\perp\beta.
\end{eqnarray}
Let
\begin{eqnarray}
\beta_\perp=\frac{\eta}{\|\eta\|}.
\end{eqnarray}
By definition,
\begin{eqnarray}
(\bra{\alpha}\otimes\bra{\eta})
\rho
(\ket{\alpha_\perp}\otimes\ket{\beta})
=
\|\eta\|^2
\ne0.
\end{eqnarray}
Taking the Hermitian conjugate gives
\begin{eqnarray}
(\bra{\alpha_\perp}\otimes\bra{\beta})
\rho
(\ket{\alpha}\otimes\ket{\eta})
=
\|\eta\|^2
\ne0.
\end{eqnarray}
When the bipartition is viewed as \(B|A\), this matrix element says that the product-null vector \(\beta\otimes\alpha\) has a nonzero tangential coherence term in the \(B\to A\) direction.  Hence the same product-null direction triggers the boundary-scaling in both steering directions.  States in the filtered standard class are therefore two-way projectively steerable.

We now show that all entangled two-qubit states of rank two are contained in this class.  Let
\begin{eqnarray}
{\rm rank}\,\rho=2.
\end{eqnarray}
Since the total Hilbert space is four-dimensional, the null space of \(\rho\) is two-dimensional.  Choose two linearly independent vectors \(\psi_1,\psi_2\) in this null space:
\begin{eqnarray}
\rho\psi_1=0,
\qquad
\rho\psi_2=0.
\end{eqnarray}
We claim that their span contains a nonzero product vector.

To see this, write a two-qubit vector as a \(2\times2\) coefficient matrix.  Namely, if
\begin{eqnarray}
\psi=
\sum_{a,b=0}^1 A_{ab}\ket a\otimes\ket b,
\end{eqnarray}
associate to it
\begin{eqnarray}
A=
\left(
\begin{array}{cc}
A_{00}&A_{01}\\
A_{10}&A_{11}
\end{array}
\right).
\end{eqnarray}
The vector \(\psi\) is a product vector if and only if
\begin{eqnarray}
\det A=0.
\end{eqnarray}
A general vector in the two-dimensional null space has the form
\begin{eqnarray}
\psi(x,y)=x\psi_1+y\psi_2,
\end{eqnarray}
with coefficient matrix
\begin{eqnarray}
A(x,y)=xA_1+yA_2.
\end{eqnarray}
Then \(\det A(x,y)\) is a homogeneous quadratic polynomial in \(x,y\).  Over the complex numbers, such a polynomial always has a nontrivial zero.  Thus one can choose
\begin{eqnarray}
(x,y)\ne(0,0)
\end{eqnarray}
such that
\begin{eqnarray}
\det A(x,y)=0.
\end{eqnarray}
The corresponding vector \(\psi(x,y)\) is a product vector and still lies in the null space of \(\rho\).  Therefore there exist nonzero local vectors \(\alpha,\beta\) such that
\begin{eqnarray}
\rho(\alpha\otimes\beta)=0.
\end{eqnarray}

Apply local unitaries to put this product-null vector in the standard position:
\begin{eqnarray}
\alpha\mapsto\ket0_A,
\qquad
\beta\mapsto\ket1_B.
\end{eqnarray}
In the new local basis,
\begin{eqnarray}
\rho\ket{01}=0.
\end{eqnarray}
Thus the \(\ket{01}\)-column of \(\rho\) is zero; by Hermiticity, the \(\ket{01}\)-row is zero as well.  In the basis
\begin{eqnarray}
\ket{00},\ket{01},\ket{10},\ket{11},
\end{eqnarray}
the state must therefore have the form
\begin{eqnarray}
\rho=
\frac{1}{\tr H}
\left(
\begin{array}{cccc}
h_{00}&0&h_{01}&h_{02}\\
0&0&0&0\\
\bar h_{01}&0&h_{11}&h_{12}\\
\bar h_{02}&0&\bar h_{12}&h_{22}
\end{array}
\right),
\label{eq:rank-two-standard-block}
\end{eqnarray}
where
\begin{eqnarray}
H=
\left(
\begin{array}{ccc}
h_{00}&h_{01}&h_{02}\\
\bar h_{01}&h_{11}&h_{12}\\
\bar h_{02}&\bar h_{12}&h_{22}
\end{array}
\right)
\end{eqnarray}
is the positive semidefinite block on
\begin{eqnarray}
{\rm span}\{\ket{00},\ket{10},\ket{11}\}.
\end{eqnarray}
Since \({\rm rank}\,\rho=2\), this block also has rank two.

From the analysis of the standard family in the previous section, if the representative in Eq.~\eqref{eq:rank-two-standard-block} had
\begin{eqnarray}
h_{02}=0,
\end{eqnarray}
then it would be PPT and hence separable.  This would contradict the assumed entanglement of \(\rho\).  Therefore any entangled rank-two state can be put, by local unitaries, into a form satisfying
\begin{eqnarray}
\rho\ket{01}=0,
\qquad
\bra{00}\rho\ket{11}\ne0.
\end{eqnarray}
This is precisely the standard boundary-scaling form.  Hence every entangled two-qubit state of rank two belongs to the filtered standard class.  Combining this with the two-way boundary-scaling argument above, we obtain
\begin{eqnarray}
{\rm rank}\,\rho=2,
\qquad
\rho\ {\rm entangled}
\quad\Longrightarrow\quad
\rho\ {\rm is\ two\mbox{-}way\ projectively\ steerable}.
\end{eqnarray}

\section{Rank-three states with a product vector spanning the null space}
We now give explicit parametrizations of the rank-three product-null-space branch.  The purpose is twofold.  First, it makes clear that the branch covered by the boundary-scaling mechanism is not a small set of examples.  Second, it gives practical forms which can be used for numerical tests or experimental constructions.

We first fix the product-null vector to be
\begin{eqnarray}
\ket{01}.
\end{eqnarray}
Thus we consider the standard representatives satisfying
\begin{eqnarray}
\rho\ket{01}=0.
\end{eqnarray}
In the computational basis
\begin{eqnarray}
\ket{00},\ket{01},\ket{10},\ket{11},
\end{eqnarray}
such a state has the form
\begin{eqnarray}
\rho_0=
\frac{1}{\tr H}
\left(
\begin{array}{cccc}
h_{00}&0&h_{01}&h_{02}\\
0&0&0&0\\
\bar h_{01}&0&h_{11}&h_{12}\\
\bar h_{02}&0&\bar h_{12}&h_{22}
\end{array}
\right),
\label{eq:rank3-standard-param-H}
\end{eqnarray}
where
\begin{eqnarray}
H=
\left(
\begin{array}{ccc}
h_{00}&h_{01}&h_{02}\\
\bar h_{01}&h_{11}&h_{12}\\
\bar h_{02}&\bar h_{12}&h_{22}
\end{array}
\right)>0.
\label{eq:rank3-H-positive-param}
\end{eqnarray}
The block \(H\) is the restriction of \(\rho_0\) to the support space
\begin{eqnarray}
{\cal S}={\rm span}\{\ket{00},\ket{10},\ket{11}\}.
\end{eqnarray}
The strict positivity of \(H\) ensures that \(\rho_0\) has rank three and that its only null direction is \(\ket{01}\).

A convenient way to make the positivity of \(H\) explicit is to use the Cholesky decomposition
\begin{eqnarray}
H=LL^\dagger,
\end{eqnarray}
with
\begin{eqnarray}
L=
\left(
\begin{array}{ccc}
a&0&0\\
x&b&0\\
z&y&c
\end{array}
\right),
\qquad
a,b,c>0,
\qquad
x,y,z\in\mathbb C.
\label{eq:rank3-cholesky-L}
\end{eqnarray}
Then
\begin{eqnarray}
H=
\left(
\begin{array}{ccc}
a^2&a\bar x&a\bar z\\
a x&|x|^2+b^2&x\bar z+b\bar y\\
a z&\bar x z+b y&|z|^2+|y|^2+c^2
\end{array}
\right)>0.
\label{eq:rank3-cholesky-H}
\end{eqnarray}
Substituting this into Eq.~\eqref{eq:rank3-standard-param-H}, we obtain the explicit family
\begin{eqnarray}
\rho_0(a,b,c,x,y,z)
=
\frac{1}{N}
\left(
\begin{array}{cccc}
a^2&0&a\bar x&a\bar z\\
0&0&0&0\\
a x&0&|x|^2+b^2&x\bar z+b\bar y\\
a z&0&\bar x z+b y&|z|^2+|y|^2+c^2
\end{array}
\right),
\label{eq:rank3-cholesky-rho}
\end{eqnarray}
where
\begin{eqnarray}
N=
a^2+b^2+c^2+|x|^2+|y|^2+|z|^2.
\end{eqnarray}
This parametrization automatically gives
\begin{eqnarray}
\rho_0\ge0,
\qquad
\tr\rho_0=1,
\qquad
{\rm rank}\,\rho_0=3,
\qquad
{\rm Ker}\,\rho_0={\rm span}\{\ket{01}\}.
\end{eqnarray}

The boundary-scaling entry is particularly simple in this parametrization.  From Eq.~\eqref{eq:rank3-cholesky-H},
\begin{eqnarray}
h_{02}=a\bar z.
\end{eqnarray}
Since \(a>0\), one has
\begin{eqnarray}
h_{02}\ne0
\quad\Longleftrightarrow\quad
z\ne0.
\end{eqnarray}
Combining this with the result of the previous sections gives
\begin{eqnarray}
z\ne0
\quad\Longleftrightarrow\quad
\rho_0(a,b,c,x,y,z)\ {\rm is\ entangled}
\Longleftrightarrow\quad
\rho_0(a,b,c,x,y,z)\ {\rm is\ two\mbox{-}way\ projectively\ steerable}.
\label{eq:rank3-cholesky-criterion}
\end{eqnarray}
The locus
\begin{eqnarray}
z=0
\end{eqnarray}
is a complex hypersurface, equivalently a real codimension-two separable boundary locus, inside this product-null-space branch.

The parameters \(a,b,c>0\) and \(x,y,z\in\mathbb C\) contain one overall positive scaling which is removed by the normalization \(N\).  Thus, after fixing the null vector to be \(\ket{01}\), the standard branch has
\begin{eqnarray}
3+6-1=8
\end{eqnarray}
real degrees of freedom.

For a general product-null vector, let
\begin{eqnarray}
\alpha
=
\cos\frac{\theta_A}{2}\ket0
+
e^{i\phi_A}\sin\frac{\theta_A}{2}\ket1,
\end{eqnarray}
and
\begin{eqnarray}
\beta
=
\cos\frac{\theta_B}{2}\ket0
+
e^{i\phi_B}\sin\frac{\theta_B}{2}\ket1,
\end{eqnarray}
with
\begin{eqnarray}
\theta_A,\theta_B\in[0,\pi],
\qquad
\phi_A,\phi_B\in[0,2\pi).
\end{eqnarray}
Choose local unitaries \(U_A,U_B\) such that
\begin{eqnarray}
U_A\ket0=\alpha,
\qquad
U_B\ket1=\beta.
\end{eqnarray}
Then the general product-null-space branch is obtained from
\begin{eqnarray}
\rho(\alpha,\beta;a,b,c,x,y,z)
=
(U_A\otimes U_B)
\rho_0(a,b,c,x,y,z)
(U_A^\dagger\otimes U_B^\dagger).
\label{eq:rank3-general-product-null-param}
\end{eqnarray}
Local unitaries do not change entanglement or projective steerability.  Hence the same criterion holds:
\begin{eqnarray}
z\ne0
\quad\Longleftrightarrow\quad
\rho(\alpha,\beta;a,b,c,x,y,z)\ {\rm is\ entangled}
\Longleftrightarrow\quad
\rho(\alpha,\beta;a,b,c,x,y,z)\ {\rm is\ two\mbox{-}way\ projectively\ steerable}.
\end{eqnarray}
The four real parameters specifying \(\alpha\otimes\beta\), together with the eight parameters in Eq.~\eqref{eq:rank3-cholesky-rho}, give a twelve-dimensional product-null-space branch.

It is also useful to give a spectral parametrization.  This form is often more convenient for numerical sampling or experimental preparation, because positivity and normalization are built in from the beginning.  Again fix the null vector to be \(\ket{01}\), so the support space is
\begin{eqnarray}
{\cal S}={\rm span}\{\ket{00},\ket{10},\ket{11}\}.
\end{eqnarray}
Let
\begin{eqnarray}
\nu_1,\nu_2,\nu_3>0,
\qquad
\nu_1+\nu_2+\nu_3=1,
\end{eqnarray}
and take a unitary matrix
\begin{eqnarray}
V\in U(3).
\end{eqnarray}
Define three orthonormal eigenvectors in \({\cal S}\) by
\begin{eqnarray}
\ket{\phi_k}
=
V_{1k}\ket{00}
+
V_{2k}\ket{10}
+
V_{3k}\ket{11},
\qquad
k=1,2,3.
\end{eqnarray}
Then
\begin{eqnarray}
\rho_0
=
\sum_{k=1}^3
\nu_k\ket{\phi_k}\bra{\phi_k}.
\label{eq:rank3-spectral-param}
\end{eqnarray}
This gives a rank-three state with
\begin{eqnarray}
\rho_0\ket{01}=0.
\end{eqnarray}
The overall phases of the eigenvectors do not affect \(\rho_0\), so the relevant unitary freedom is
\begin{eqnarray}
V\in U(3)/U(1)^3,
\end{eqnarray}
which has six real degrees of freedom.  Together with the two independent spectral weights, this again gives eight real parameters.

In the spectral parametrization, the key matrix element is
\begin{eqnarray}
\bra{00}\rho_0\ket{11}
=
\sum_{k=1}^3
\nu_k V_{1k}\overline{V_{3k}}.
\label{eq:rank3-spectral-coherence}
\end{eqnarray}
Therefore
\begin{eqnarray}
\sum_{k=1}^3
\nu_k V_{1k}\overline{V_{3k}}
\ne0
\quad\Longleftrightarrow\quad
\rho_0\ {\rm is\ entangled}
\Longleftrightarrow\quad
\rho_0\ {\rm is\ two\mbox{-}way\ projectively\ steerable}.
\label{eq:rank3-spectral-criterion}
\end{eqnarray}
Thus, in this representation, the condition is simply that the mixture retains a nonzero \(\ket{00}\leftrightarrow\ket{11}\) coherence.

For explicit scans one may write a generic chart of \(U(3)/U(1)^3\) as a product of three complex plane rotations:
\begin{eqnarray}
V=
R_{23}(\theta_{23},\varphi_{23})
R_{13}(\theta_{13},\varphi_{13})
R_{12}(\theta_{12},\varphi_{12}),
\end{eqnarray}
where
\begin{eqnarray}
\theta_{ij}\in[0,\pi/2],
\qquad
\varphi_{ij}\in[0,2\pi).
\end{eqnarray}
For example,
\begin{eqnarray}
R_{12}(\theta,\varphi)
=
\left(
\begin{array}{ccc}
\cos\theta&e^{i\varphi}\sin\theta&0\\
-e^{-i\varphi}\sin\theta&\cos\theta&0\\
0&0&1
\end{array}
\right),
\end{eqnarray}
and \(R_{13}\), \(R_{23}\) are defined analogously on the corresponding two-dimensional coordinate planes.  In this chart, the spectral data
\begin{eqnarray}
\nu_1,\nu_2,\nu_3;
\qquad
\theta_{12},\theta_{13},\theta_{23};
\qquad
\varphi_{12},\varphi_{13},\varphi_{23}
\end{eqnarray}
give an explicit eight-parameter description after imposing \(\nu_1+\nu_2+\nu_3=1\).  The quantity to monitor is
\begin{eqnarray}
C_{03}
=
\sum_{k=1}^3
\nu_k V_{1k}\overline{V_{3k}}.
\end{eqnarray}
The condition \(C_{03}\ne0\) is exactly the entanglement and two-way projective-steering condition on this branch.

Several lower-dimensional sections of the above parametrizations are useful for illustration.  Setting
\begin{eqnarray}
x=0,
\qquad
y=0
\end{eqnarray}
in Eq.~\eqref{eq:rank3-cholesky-rho} gives an \(X\)-type family,
\begin{eqnarray}
\rho_X(a,b,c,z)
=
\frac{1}{a^2+b^2+c^2+|z|^2}
\left(
\begin{array}{cccc}
a^2&0&0&a\bar z\\
0&0&0&0\\
0&0&b^2&0\\
a z&0&0&|z|^2+c^2
\end{array}
\right).
\label{eq:rank3-X-family}
\end{eqnarray}
Here again \(z\ne0\) is equivalent to entanglement and to two-way projective steerability.

A more physical section is obtained by mixing two Bell states with a product state:
\begin{eqnarray}
\rho_{p,q,r}^{(\Phi)}
=
p\ket{\Phi^+}\bra{\Phi^+}
+
q\ket{\Phi^-}\bra{\Phi^-}
+
r\ket{10}\bra{10},
\label{eq:rank3-bellpair-product-family}
\end{eqnarray}
where
\begin{eqnarray}
p,q,r>0,
\qquad
p+q+r=1,
\end{eqnarray}
and
\begin{eqnarray}
\ket{\Phi^\pm}
=
\frac{\ket{00}\pm\ket{11}}{\sqrt2}.
\end{eqnarray}
In the computational basis,
\begin{eqnarray}
\rho_{p,q,r}^{(\Phi)}
=
\left(
\begin{array}{cccc}
\frac{p+q}{2}&0&0&\frac{p-q}{2}\\
0&0&0&0\\
0&0&r&0\\
\frac{p-q}{2}&0&0&\frac{p+q}{2}
\end{array}
\right).
\end{eqnarray}
Thus
\begin{eqnarray}
\bra{00}\rho_{p,q,r}^{(\Phi)}\ket{11}
=
\frac{p-q}{2}.
\end{eqnarray}
Consequently,
\begin{eqnarray}
p\ne q
\quad\Longleftrightarrow\quad
\rho_{p,q,r}^{(\Phi)}\ {\rm is\ entangled}
\Longleftrightarrow\quad
\rho_{p,q,r}^{(\Phi)}\ {\rm is\ two\mbox{-}way\ projectively\ steerable}.
\end{eqnarray}
When \(p=q\), the two Bell components cancel the \(\ket{00}\leftrightarrow\ket{11}\) coherence and the state lies on the separable boundary of the branch.

Another simple section is a mixture of two product states and one entangled pure state:
\begin{eqnarray}
\rho_{p,q,r,\theta,\varphi}
=
p\ket{00}\bra{00}
+
q\ket{10}\bra{10}
+
r\ket{\phi_{\theta,\varphi}}\bra{\phi_{\theta,\varphi}},
\label{eq:rank3-two-products-one-entangled}
\end{eqnarray}
where
\begin{eqnarray}
p,q,r>0,
\qquad
p+q+r=1,
\end{eqnarray}
and
\begin{eqnarray}
\ket{\phi_{\theta,\varphi}}
=
\cos\theta\,\ket{00}
+
e^{i\varphi}\sin\theta\,\ket{11},
\qquad
0<\theta<\frac{\pi}{2}.
\end{eqnarray}
The relevant matrix element is
\begin{eqnarray}
\bra{00}\rho_{p,q,r,\theta,\varphi}\ket{11}
=
r e^{-i\varphi}\cos\theta\sin\theta,
\end{eqnarray}
which is nonzero for the stated parameter range.  Hence the whole family is entangled and two-way projectively steerable.

Finally, the standard form may be hidden by an invertible filter on Bob's side.  For any standard representative \(\rho_0\) and any invertible matrix \(G\), define
\begin{eqnarray}
\rho_G
=
\frac{
(I\otimes G)\rho_0(I\otimes G^\dagger)
}{
\tr[(I\otimes G)\rho_0(I\otimes G^\dagger)]
}.
\end{eqnarray}
Then \(\rho_G\) remains in the filtered standard class.  Its null vector is
\begin{eqnarray}
\ket0\otimes (G^\dagger)^{-1}\ket1.
\end{eqnarray}
Thus the zero row and column may no longer be visible in the computational basis, but the product-null-space structure and the boundary-scaling mechanism are unchanged.

The Cholesky parametrization gives the most direct theoretical criterion, namely \(z\ne0\).  The spectral parametrization gives a convenient preparation-oriented description, where the same condition becomes the nonvanishing of the support coherence \(C_{03}\) in Eq.~\eqref{eq:rank3-spectral-coherence}.  Both parametrizations describe the same rank-three product-null-space branch.

\section{Higher-dimensional and multipartite support-kernel criteria}

The boundary-contact obstruction established above is two-dimensional, but the essential mechanism is local.  It takes place in the trusted slice selected by one support vector and one kernel vector.  Thus the relevant geometry is not the global geometry of the Bloch ball, but the local boundary geometry of the trusted conditional state space.  In the two-qubit case, a boundary contact of Bob's conditional state is necessarily a contact with a pure point of the Bloch sphere.  In higher dimensions, the trusted conditional state may touch a general rank-deficient point of the density-matrix cone.  The appropriate formulation is therefore a support-kernel criterion for density matrices.

We formulate the result for a general steering cut
\begin{eqnarray}
X|Y,
\end{eqnarray}
where \(X\) is the untrusted measuring side and \(Y\) is the trusted steered side.  The trusted side may be a single high-dimensional system, or a multipartite joint system, for instance
\begin{eqnarray}
{\cal H}_Y={\cal H}_{B_1}\otimes\cdots\otimes{\cal H}_{B_n}.
\end{eqnarray}
Thus steering from \(X\) to \(B_1\cdots B_n\) can first be regarded as ordinary bipartite steering with a high-dimensional trusted system.

Assume that \(\dim {\cal H}_X\ge2\), and choose two orthonormal vectors on the untrusted side,
\begin{eqnarray}
\ket{\alpha_0},
\qquad
\ket{\alpha_1}.
\end{eqnarray}
We write the corresponding \(2\times2\) block of \(\rho\), with respect to this two-dimensional subspace of \(X\), as
\begin{eqnarray}
\rho_{01}
=
\left(
\begin{array}{cc}
A&B\\
B^\dagger&D
\end{array}
\right)_X,
\label{eq:highdim-block}
\end{eqnarray}
where
\begin{eqnarray}
A=\bra{\alpha_0}\rho\ket{\alpha_0},
\qquad
B=\bra{\alpha_0}\rho\ket{\alpha_1},
\qquad
D=\bra{\alpha_1}\rho\ket{\alpha_1}
\end{eqnarray}
are operators on \({\cal H}_Y\).  If \({\cal H}_X\) has dimension larger than two, \(\rho\) may also have blocks involving vectors orthogonal to \(\operatorname{span}\{\ket{\alpha_0},\ket{\alpha_1}\}\).  These blocks do not contribute to the one-parameter projective family used below.

Consider the rank-one projective outcome
\begin{eqnarray}
\ket{\xi_t}
=
\frac{\ket{\alpha_0}+t\ket{\alpha_1}}{\sqrt{1+t^2}},
\qquad
P_t=\ket{\xi_t}\bra{\xi_t}.
\end{eqnarray}
The corresponding unnormalized conditional state on the trusted side is
\begin{eqnarray}
\sigma_t
&=&
\operatorname{tr}_X[(P_t\otimes I_Y)\rho]
\nonumber\\
&=&
\frac{A+t(B+B^\dagger)+t^2D}{1+t^2}.
\label{eq:highdim-sigma-t}
\end{eqnarray}

Suppose first that
\begin{eqnarray}
A\neq0,
\qquad
A\ge0,
\qquad
{\rm Ker}\,A\neq\{0\}.
\label{eq:highdim-boundary-contact}
\end{eqnarray}
Then \(\sigma_0=A\) lies on the boundary of the density-matrix cone on \({\cal H}_Y\).  We shall say that this boundary contact has support-kernel first-order coherence if
\begin{eqnarray}
P_{\supp A}BP_{\Ker A}\neq0.
\label{eq:highdim-support-kernel-coupling}
\end{eqnarray}
Equivalently, there exist
\begin{eqnarray}
\ket{\phi}\in{\rm supp}A,
\qquad
\ket{\beta}\in{\rm Ker}A
\end{eqnarray}
such that
\begin{eqnarray}
v=\bra{\phi}B\ket{\beta}\neq0.
\label{eq:highdim-v-nonzero}
\end{eqnarray}
Let
\begin{eqnarray}
{\cal K}=\operatorname{span}\{\ket{\phi},\ket{\beta}\}
\end{eqnarray}
and let \(\Pi_{\cal K}\) be the projector onto \({\cal K}\).  Compressing Eq.~\eqref{eq:highdim-sigma-t} to \({\cal K}\), and using the ordered basis
\begin{eqnarray}
\ket{\phi},
\qquad
\ket{\beta},
\end{eqnarray}
we obtain
\begin{eqnarray}
\Pi_{\cal K}\sigma_t\Pi_{\cal K}
=
\left(
\begin{array}{cc}
a+O(t)&tv+O(t^2)\\
t\bar v+O(t^2)&t^2d+O(t^3)
\end{array}
\right),
\label{eq:highdim-compressed-scaling}
\end{eqnarray}
where
\begin{eqnarray}
a=\bra{\phi}A\ket{\phi}>0,
\qquad
v\neq0,
\qquad
d=\bra{\beta}D\ket{\beta}\ge0.
\end{eqnarray}
Thus the same local scaling pattern as in the two-qubit proof appears inside a trusted two-dimensional subspace:
\begin{eqnarray}
|b_t|\sim t,
\qquad
 d_t=O(t^2).
\label{eq:highdim-local-scaling}
\end{eqnarray}
Here the notation \(d_t=O(t^2)\) is deliberate.  If \(d=0\), the lower-right entry may vanish to higher order; the boundary-scaling obstruction only requires an upper bound of order \(t^2\).

Let us explain why the lower-right entry in Eq.~\eqref{eq:highdim-compressed-scaling} has no first-order term.  Since \(\ket{\beta}\in{\rm Ker}A\),
\begin{eqnarray}
\bra{\alpha_0,\beta}\rho\ket{\alpha_0,\beta}
=
\bra{\beta}A\ket{\beta}
=
0.
\end{eqnarray}
As \(\rho\ge0\), this implies
\begin{eqnarray}
\rho\ket{\alpha_0,\beta}=0.
\end{eqnarray}
In particular,
\begin{eqnarray}
B^\dagger\ket{\beta}=0.
\end{eqnarray}
Therefore
\begin{eqnarray}
\bra{\beta}(B+B^\dagger)\ket{\beta}=0,
\end{eqnarray}
and the \(\ket{\beta}\)-diagonal entry of \(\sigma_t\) starts at order \(t^2\).

We now show that this local two-dimensional scaling already rules out an LHS model for the original \(X\to Y\) projective assemblage.  Suppose, for contradiction, that the original assemblage had an LHS representation
\begin{eqnarray}
\sigma_t
=
\int f_t(\lambda)\tau_\lambda\,d\mu(\lambda),
\qquad
0\le f_t(\lambda)\le1,
\label{eq:highdim-lhs}
\end{eqnarray}
where \(\mu\) is a finite hidden-variable measure and each \(\tau_\lambda\) is a normalized state on \({\cal H}_Y\).  Compressing by \(\Pi_{\cal K}\) gives
\begin{eqnarray}
\widetilde\sigma_t
:=
\Pi_{\cal K}\sigma_t\Pi_{\cal K}
=
\int f_t(\lambda)\Pi_{\cal K}\tau_\lambda\Pi_{\cal K}\,d\mu(\lambda).
\end{eqnarray}
For those \(\lambda\) with
\begin{eqnarray}
r_\lambda
=
\operatorname{tr}(\Pi_{\cal K}\tau_\lambda\Pi_{\cal K})>0,
\end{eqnarray}
define
\begin{eqnarray}
\widetilde\tau_\lambda
=
\frac{\Pi_{\cal K}\tau_\lambda\Pi_{\cal K}}{r_\lambda},
\qquad
 d\widetilde\mu(\lambda)=r_\lambda\,d\mu(\lambda).
\end{eqnarray}
The points with \(r_\lambda=0\) do not contribute to the compressed assemblage and may be discarded.  Then
\begin{eqnarray}
\widetilde\sigma_t
=
\int f_t(\lambda)\widetilde\tau_\lambda\,d\widetilde\mu(\lambda),
\label{eq:compressed-lhs}
\end{eqnarray}
which is again an LHS model with a finite hidden-state measure, now on the two-dimensional trusted space \({\cal K}\).

However, Eq.~\eqref{eq:highdim-compressed-scaling} has the two-dimensional boundary-contact scaling form
\begin{eqnarray}
|b_t|\ge lt,
\qquad
 d_t\le ct^2
\end{eqnarray}
for sufficiently small \(t>0\).  By the two-dimensional boundary-contact scaling obstruction, no finite-measure LHS model can simulate this compressed family.  This contradicts Eq.~\eqref{eq:compressed-lhs}.  Hence the original projective assemblage is not LHS, and therefore
\begin{eqnarray}
\rho\ \text{is projectively steerable from }X\text{ to }Y.
\label{eq:highdim-steering-conclusion}
\end{eqnarray}

The same support-kernel coherence also implies NPT entanglement across the cut \(X|Y\).  Choose an orthonormal basis of \({\cal H}_Y\) containing \(\ket{\phi}\) and \(\ket{\beta}\), and take the partial transpose \(\rho^{\Gamma_Y}\) with respect to this trusted-side basis.  Restrict \(\rho^{\Gamma_Y}\) to the two-dimensional subspace
\begin{eqnarray}
\operatorname{span}\{
\ket{\alpha_0,\beta},
\ket{\alpha_1,\phi}
\}.
\end{eqnarray}
The first diagonal entry is
\begin{eqnarray}
\bra{\alpha_0,\beta}\rho^{\Gamma_Y}\ket{\alpha_0,\beta}
=
\bra{\beta}A\ket{\beta}
=
0.
\end{eqnarray}
The corresponding off-diagonal entry is
\begin{eqnarray}
\bra{\alpha_0,\beta}\rho^{\Gamma_Y}\ket{\alpha_1,\phi}
&=&
\bra{\alpha_0,\phi}\rho\ket{\alpha_1,\beta}
\nonumber\\
&=&
\bra{\phi}B\ket{\beta}
=
v\neq0.
\end{eqnarray}
Thus this principal \(2\times2\) submatrix has the form
\begin{eqnarray}
\left(
\begin{array}{cc}
0&v\\
\bar v&*
\end{array}
\right),
\end{eqnarray}
whose determinant is
\begin{eqnarray}
-|v|^2<0.
\end{eqnarray}
Therefore
\begin{eqnarray}
\rho^{\Gamma_Y}\not\ge0.
\end{eqnarray}
Consequently, \(\rho\) is NPT, and hence entangled, across the cut \(X|Y\).

The preceding argument proves the following higher-dimensional boundary-scaling criterion.

\begin{proposition}[Higher-dimensional support-kernel criterion]
Let \(\rho\) be a state on \({\cal H}_X\otimes{\cal H}_Y\), and let \(\ket{\alpha_0},\ket{\alpha_1}\in{\cal H}_X\) be orthonormal.  Write the corresponding \(2\times2\) block of \(\rho\) as
\begin{eqnarray}
\rho_{01}
=
\left(
\begin{array}{cc}
A&B\\
B^\dagger&D
\end{array}
\right)_X.
\end{eqnarray}
Assume that
\begin{eqnarray}
A\neq0,
\qquad
{\rm Ker}\,A\neq\{0\},
\qquad
P_{\supp A}BP_{\Ker A}\neq0.
\end{eqnarray}
Then \(\rho\) is NPT across the cut \(X|Y\), and the projective assemblage generated by measurements on \(X\) is not LHS.  Hence \(\rho\) is projectively steerable from \(X\) to \(Y\).
\label{prop:highdim-boundary-scaling}
\end{proposition}

Equivalently,
\begin{eqnarray}
\left.
\begin{array}{l}
A\neq0,\\[1mm]
{\rm Ker}\,A\neq\{0\},\\[1mm]
P_{\supp A}BP_{\Ker A}\neq0
\end{array}
\right\}
&\Longrightarrow&
\left\{
\begin{array}{l}
\rho^{\Gamma_Y}\not\ge0\quad{\rm across\ the\ cut}\ X|Y,\\[1mm]
\rho\ {\rm is\ projectively\ steerable\ from}\ X\ {\rm to}\ Y.
\end{array}
\right.
\label{eq:highdim-main-criterion}
\end{eqnarray}
When specialized to the two-qubit product-null family, this criterion becomes the single-entry condition \(h_{02}\neq0\).  In the general fixed-cut form, the trusted conditional state touches the boundary because \(A\) is rank-deficient, and the first-order term \(B\) has a nonzero component from the kernel of \(A\) into its support.

The conclusion of Proposition~\ref{prop:highdim-boundary-scaling} is stronger than entanglement across the same cut.  Projective steering implies entanglement, but the converse is false in general.  For example, take a two-qubit Werner state \(\rho_W(v)\) in a parameter range where it is entangled but unsteerable for projective measurements, and tensor it with an arbitrary spectator state
\begin{eqnarray}
\tau_{B_2\cdots B_n}.
\end{eqnarray}
Then
\begin{eqnarray}
\rho_W(v)_{AB_1}\otimes\tau_{B_2\cdots B_n}
\end{eqnarray}
remains entangled across the cut
\begin{eqnarray}
A|B_1\cdots B_n,
\end{eqnarray}
but the LHS model of the Werner state extends trivially to the spectator system.  Hence it is not projectively steerable from \(A\) to the joint trusted system \(B_1\cdots B_n\).  Thus, in higher-dimensional and multipartite settings,
\begin{eqnarray}
{\rm entanglement}
\not\Longrightarrow
{\rm projective\ steering}
\end{eqnarray}
in general.

It is also important that, beyond the two-qubit case, PPT is not equivalent to separability.  In general,
\begin{eqnarray}
\rho^{\Gamma_Y}\not\ge0
\Longrightarrow
\rho\ \text{is entangled},
\end{eqnarray}
but the converse may fail because PPT entangled states can exist.  Hence Proposition~\ref{prop:highdim-boundary-scaling} should be read as a strong sufficient mechanism,
\begin{eqnarray}
\text{boundary contact plus support-kernel coherence}
\Longrightarrow
\text{NPT entanglement and projective steering},
\end{eqnarray}
not as a complete characterization of higher-dimensional entanglement or steering.

There is, however, an important pure-contact situation in which the equivalence structure survives in arbitrary finite dimensions.  Suppose that the boundary contact is a pure-state contact,
\begin{eqnarray}
A=a\ket{\phi}\bra{\phi},
\qquad
 a>0.
\label{eq:highdim-pure-contact}
\end{eqnarray}
Then \({\rm supp}A=\operatorname{span}\{\ket{\phi}\}\).  Positivity of the block matrix in Eq.~\eqref{eq:highdim-block} forces the range of \(B\) to lie inside \({\rm supp}A\).  Indeed, for every \(\ket{\beta}\in{\rm Ker}A\), positivity gives
\begin{eqnarray}
\rho_{01}\ket{\alpha_0,\beta}=0,
\end{eqnarray}
and hence
\begin{eqnarray}
B^\dagger\ket{\beta}=0.
\end{eqnarray}
Equivalently, \(B\) has range contained in \(\operatorname{span}\{\ket{\phi}\}\).  Therefore
\begin{eqnarray}
B=\ket{\phi}\bra{w}
\end{eqnarray}
for some vector \(\ket{w}\in{\cal H}_Y\).  Decompose
\begin{eqnarray}
\ket{w}=w_0\ket{\phi}+\ket{w_\perp},
\qquad
\ket{w_\perp}\in{\rm Ker}A.
\end{eqnarray}
Then
\begin{eqnarray}
P_{\supp A}BP_{\Ker A}\neq0
\quad\Longleftrightarrow\quad
\ket{w_\perp}\neq0.
\end{eqnarray}
If \(\ket{w_\perp}\neq0\), Proposition~\ref{prop:highdim-boundary-scaling} proves NPT entanglement and projective steering.

Conversely, suppose that
\begin{eqnarray}
P_{\supp A}BP_{\Ker A}=0.
\end{eqnarray}
Then \(\ket{w_\perp}=0\), and hence there exists a scalar \(\kappa\) such that
\begin{eqnarray}
B=\kappa\ket{\phi}\bra{\phi}.
\end{eqnarray}
The Schur complement gives
\begin{eqnarray}
D-\frac{|\kappa|^2}{a}\ket{\phi}\bra{\phi}\ge0.
\label{eq:pure-contact-schur}
\end{eqnarray}
Consequently, the block state \(\rho_{01}\) can be decomposed as
\begin{eqnarray}
\rho_{01}
&=&
\ket{\eta}\bra{\eta}_X\otimes\ket{\phi}\bra{\phi}_Y
\nonumber\\
&&+
\ket{\alpha_1}\bra{\alpha_1}_X
\otimes
\left(
D-\frac{|\kappa|^2}{a}\ket{\phi}\bra{\phi}
\right),
\label{eq:pure-contact-separable-decomposition}
\end{eqnarray}
where
\begin{eqnarray}
\ket{\eta}
=
\sqrt a\,\ket{\alpha_0}
+
\frac{\bar\kappa}{\sqrt a}\ket{\alpha_1}.
\end{eqnarray}
Thus the two-dimensional \(X\)-block \(\rho_{01}\) is separable across \(X|Y\).

Therefore, for the pure-contact standard family supported on
\begin{eqnarray}
\operatorname{span}\{\ket{\alpha_0},\ket{\alpha_1}\}\otimes{\cal H}_Y,
\end{eqnarray}
one recovers the same equivalence structure as in the two-qubit product-null-space boundary:
\begin{eqnarray}
{\rm entanglement}
\quad\Longleftrightarrow\quad
{\rm NPT}
\quad\Longleftrightarrow\quad
P_{\supp A}BP_{\Ker A}\neq0
\quad\Longleftrightarrow\quad
{\rm projective\ steering}.
\label{eq:pure-contact-equivalence}
\end{eqnarray}
If additional blocks outside \(\operatorname{span}\{\ket{\alpha_0},\ket{\alpha_1}\}\) are present on the untrusted side, the absence of support-kernel coherence in this particular block only rules out this boundary-scaling mechanism.  It does not by itself imply separability of the full state.

For a general mixed boundary contact, namely \({\rm rank}A>1\), the equivalence in Eq.~\eqref{eq:pure-contact-equivalence} no longer holds in general.  The support-kernel coupling condition should then be viewed as a sufficient condition for NPT entanglement and steering, not as an exhaustive characterization of entanglement.

Finally, in multipartite applications one must distinguish steering across a fixed cut from genuine multipartite steering.  The argument above proves projective steering for a specified cut
\begin{eqnarray}
X|Y.
\end{eqnarray}
When \(Y=B_1\cdots B_n\), the trusted side is treated as a single joint system.  This does not automatically imply genuine multipartite steering, which would require excluding all hybrid LHS models associated with different bipartitions.  Such stronger multipartite conclusions would require additional boundary-scaling obstructions across the relevant cuts.